\documentclass[12pt]{iopart}
\usepackage{upgreek}
\usepackage{graphicx}
\usepackage{subcaption}
\usepackage[normalem]{ulem}

\usepackage{color}
\definecolor{brickred}{rgb}{0.8, 0.25, 0.33} % http://latexcolor.com/

\usepackage{graphicx}% Include figure files

\usepackage{bm}% bold math
\usepackage{times}
\usepackage{cite}
\usepackage[colorlinks,
            citecolor=red,
            urlcolor=blue,
            bookmarks=false,
            hypertexnames=true]{hyperref}

\usepackage{hyperref}
\hypersetup{colorlinks=true}

\begin{document}

%\setulcolor{red} set underlining color
%\setstcolor{green} set overstriking color
%\sethlcolor{yellow} set highlighting color
%\so{letterspacing} \
%\caps{CAPITALS, Small Capitals}\
%\ul{underlining}\
%\st{overstriking} \
%\hl{highlighting} 

% \preprint{APS/123-QED}

\title[]{Diamond-shaped superconducting interference patterns in NbTiN weak-link Josephson junctions under in-plane magnetic fields}% Force line breaks with \\

\author{Kui Zhao$^{1,2,\dagger}$,
        Jianfei Xiao$^{1,3,\dagger}$,
        Huaiyuan Liu$^{1,\dagger}$,
        Linfeng Tu$^{1,3}$,
        Yiwen Ma$^{1,3}$,
        Jiangbo He$^{1,3}$,
        Mingli Liu$^{1,3}$,
        Ruiyang Jiang$^{1,3}$,
        Zhongmou Jia$^{1,3}$,
        Shang Zhu$^{1,3}$,
        Yunteng Shi$^{1,3}$,
        Zhaozheng Lyu$^{1,3}$,
        Jie Shen$^{1,3,4}$,
        Guangtong Liu$^{1,3,4}$,
        Li Lu$^{1,3,4,*}$,
        Fanming Qu$^{1,3,4,*}$
}
\address{$^{*}$ \textit{Corresponding author}}
\address{$^{1}$ Beijing National Laboratory for Condensed Matter Physics, Institute of Physics, Chinese Academy of Sciences, Beijing 100190, China.}
\address{$^{2}$ Beijing Key Laboratory of Fault-Tolerant Quantum Computing, Beijing Academy of Quantum Information Sciences, Beijing 100193, China.}
\address{$^{3}$ School of Physical Sciences, University of Chinese Academy of Sciences, Beijing 100049, China.}
\address{$^{4}$ Songshan Lake Materials Laboratory, Dongguan, Guangdong 523808, China.}

\address{$^{\dagger}$ These authors contributed equally to this work.}

\ead{fanmingqu@iphy.ac.cn}
\ead{lilu@iphy.ac.cn}

\vspace{10pt}
\begin{indented}
\item[]
\end{indented}


\date{\today}% It is always \today, today,

\begin{abstract}

The application of in-plane magnetic fields to Josephson junctions enables fundamental exploration of quantum phenomena, including Zeeman-driven 0-$\pi$ transitions and planar topological superconductivity. However, intrinsic orbital effects arising from nanoscale rippled geometries in practical devices can dominate phase interference signatures, complicating their interpretation. Here, we experimentally probe superconducting interference in NbTiN weak-link Josephson junctions under combined perpendicular and in-plane magnetic fields. The critical supercurrent reveals a distinct diamond-shaped interference pattern, with nodes progressively opening and evolving into V-shaped features, reminiscent of suppression-recovery patterns associated with 0-$\pi$ transitions. We theoretically analyze the interplay between orbital effects from rippled geometries and non-uniform supercurrent density distributions, demonstrating that their synergistic interaction could reproduce the experimentally observed interference evolution. Our findings elucidate how geometric imperfections and current inhomogeneity cooperatively reshape phase interference, providing critical insights into orbital-dominated phenomena in Josephson systems. 

\end{abstract}


%\tableofcontents

\section{Introduction}

Josephson junctions—comprising two superconducting electrodes coupled via a weak link that can be a tunnel barrier, normal metal, or constriction—sustain dissipationless supercurrents governed by the phase difference between the electrodes. The integration of Zeeman effects into these systems has emerged as a pivotal avenue for exploring exotic quantum phenomena, such as Fulde–Ferrell–Larkin–Ovchinnikov (FFLO) states~\cite{FFLO1PhysRev.135.A550,FFLO2Quasiclassical}, anomalous Josephson effects~\cite{AnomalousJosephsoneffect2014,AnomalousJosephsoneffectQD}, planar topological superconductivity~\cite{PhysRevLett.118.107701,PhysRevX.7.021032,Nature569,Nature5691}, and the Josephson diode effect~\cite{UniversalJosephsondiodeeffect,Josephsondiodeeffectsemimetal,Superconductingdiodeeffect,field-freeJosephsondiode}. These phenomena hold promise for applications in superconducting spintronics~\cite{Superconductingspintronicsnphys3242}, quantum computing ~\cite{NaturePhys6593,JournalofAppliedPhysics94}, and cryogenic memory ~\cite{cryogenicmemory}.

The Zeeman effect induces spin-polarized Fermi surfaces, enabling Cooper pairs with finite center-of-mass momentum to persist. This momentum manifests as a spatially oscillatory component in the superconducting order parameter~\cite{FFLORevModPhys.77.935}, driving transitions between ground states with 0 and $\pi$ phase differences—so-called 0-$\pi$ transitions. While such transitions have been extensively studied in ferromagnetic Josephson junctions~\cite{PhysRevLett.89.137007,PhysRevLett.90.167001,PhysRevLett.89.187004,PhysRevLett.96.197003,PhysRevB.73.174506,PhysRevLett.97.177003}, recent experiments have demonstrated their realization in a variety of systems, including semiconductors ~\cite{ControlledfinitemomentumpairingHgTequantumwells,BallisticsuperconductivityInSbquantumwells}, semimetals~\cite{ZeemanEffectInducedTransitionsDiracSemimetal,4piperiodicAndreevboundstatesinaDiracsemimetalNatureMater}, and topological insulators~\cite{FinitemomentumCooperpairing}, where suppression-revival patterns of the supercurrent are attributed to Zeeman-driven 0-$\pi$ transitions. However, a complication arises in realistic devices: intrinsic rippled structures in two-dimensional materials~\cite{fasolino_intrinsic_2007,bao_controlled_2009,ludacka_situ_2018,yu_atomscopic_2024,ng_improving_2022,Strain_engineering_2021} generate orbital effects via out-of-plane magnetic field components, which can mimic 0-$\pi$ transition signatures~\cite{CombinedZeemanandorbitaleffect,PlanargrapheneNbSeJosephsonjunctionsPhysRevB.103.115401}. This interplay necessitates rigorous discrimination between orbital and Zeeman-dominated mechanisms in phase-coherent transport under in-plane magnetic fields.

In this work, we investigate Josephson junctions comprising NbTiN weak links, employing Bi$_{2}$O$_{2}$Se flakes as a structural support. The weak link naturally forms at the edge of the Bi$_{2}$O$_{2}$Se flake due to the height disparity between the flake and the underlying Si/SiO$_2$ substrate. These junctions exhibit conventional superconducting transport characteristics, including quantum interference under perpendicular magnetic fields and Shapiro-step responses to microwave irradiation~\cite{supp}. Notably, however, the magnetic interference pattern deviates from the standard Fraunhofer-like behavior, instead resembling that of a superconducting quantum interference device (SQUID), a signature of spatially non-uniform current flow within the junction~\cite{grapheneNbTiNjunctionsPhysRevB.85.205404,Edgemodesuperconductivity,InducedsuperconductivityinthequantumspinHalledge,SQUIDpatterndisruptionPhysRevB.102.165407,Bi2O2SeBasedJosephsonJunction,Spatiallyresolvededgecurrentsandguidedwaveelectronicstatesingraphene,EdgesuperconductivityinmultilayerWTe2Josephsonjunction}. When subjected to an in-plane magnetic field, the interference pattern undergoes a striking transformation: individual nodes progressively split and elongate into V-shaped minima, collectively forming a diamond-like diagram. To elucidate this phenomenon, we model the interplay between orbital effects—arising from nanoscale ripples inherent to the non-ideal planar geometry of the weak link—and inhomogeneous current distributions. Our theoretical framework reveals that the observed diamond-shaped pattern emerges from the synergistic interplay of ripple-induced orbital phase modulation and spatially varying current density. This mechanism reproduces the experimental node evolution with remarkable fidelity, underscoring the critical role of geometric imperfections in shaping phase-coherent transport under in-plane magnetic fields.

\section{Transport characteristics}

\begin{figure}[htbp]
\centering
\includegraphics[width=0.65\textwidth]{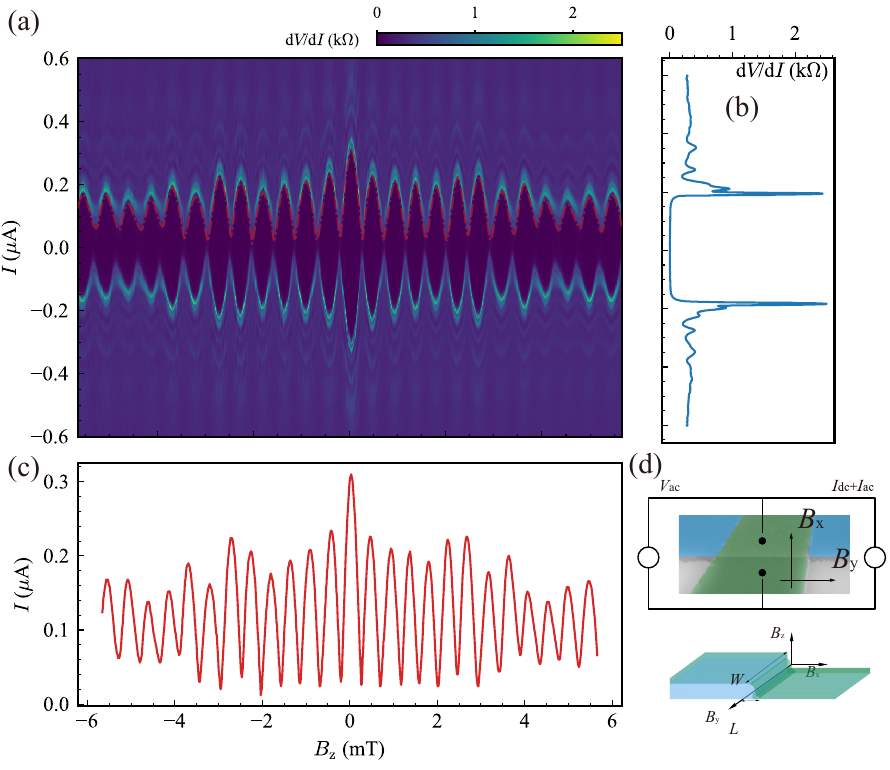}% Here is how to import EPS art
\caption{\label{fig:Fig_1} (a) Two-dimensional map of differential resistance (d$V$/d$I$) as a function of DC current bias ($I$) and perpendicular magnetic flux density ($B_z$) measured at $T$ = 10 mK, exhibiting SQUID-like interference modulations. The critical supercurrent $I_c$ (red curve) is defined by the differential resistance threshold (82.3 $\Omega$) demarcating the superconducting regime. (b) The differential resistance trace as a function of $I$ at $B_z$ = 0 is presented. (c) The extracted critical supercurrent $I_c$ (the same red curve as in (a)) is shown. (d) False-color scanning electron microscopy image (top) of Josephson junction (JJA) and 3D schematic is provided to illustrating the self-formed weak-link junction geometry. The NbTiN superconducting electrode (green, approximately $65$ nm thick) spans the edge of the Bi$_2$O$_2$Se flake (blue, approximately $30$ nm thick), where the weak-link junction is formed.}
\end{figure}

Josephson junctions with NbTiN weak links were fabricated via standard electron-beam lithography, utilizing Bi$_{2}$O$_{2}$Se flakes as a structural substrate. A 65-nm-thick NbTiN film was sputtered onto Bi$_{2}$O$_{2}$Se flakes ($\sim$30 nm thick), with the weak link self-forming at the flake edge due to the height disparity between the flake and the Si/SiO$_2$substrate. Devices were then wire-bonded and measured in a cryo-free $^3$He/$^4$He dilution refrigerator equipped with a three-axis vector magnet and reaching a base temperature of $T$ = 10 mK. The transport measurements were carried out in a quasi-four-terminal configuration [Fig.~\ref{fig:Fig_1}(d)] using standard lock-in techniques .

We focus on Josephson junction A (JJA) in the main text~\cite{supp}. Figure~\ref{fig:Fig_1}(a) displays the differential resistance d$V$/d$I$ of JJA as a function of DC bias current, revealing a critical supercurrent $I_c\cong$ 275 nA through the characteristic switching between superconducting and resistive states. Next, we present the superconducting interference results when applying a magnetic field $B_z$ perpendicular to the junction plane. Figure~\ref{fig:Fig_1}(b)  shows d$V$/d$I$ as a function of $B_z$ and $I$. Instead of the conventional Fraunhofer-like pattern, a SQUID-like interference pattern emerges, signaling inhomogeneous current distribution in the junction~\cite{grapheneNbTiNjunctionsPhysRevB.85.205404,Edgemodesuperconductivity,InducedsuperconductivityinthequantumspinHalledge,SQUIDpatterndisruptionPhysRevB.102.165407,Bi2O2SeBasedJosephsonJunction,Spatiallyresolvededgecurrentsandguidedwaveelectronicstatesingraphene,EdgesuperconductivityinmultilayerWTe2Josephsonjunction}. The apparent period is $\sim$0.5 mT corresponding to an effective area 4 $\mathit{\mu} m^2$. Based on the expected period $\Phi_0 / [(L_{eff}+2\lambda_L)W]$ with flux quantum $\Phi_0 = h / 2 e$, the geometric width of the junction $W$  $\sim$ 4 $\mathit{\mu}$m, and the London penetration length $\lambda_L$ = 350 nm~\cite{AbrikosovvortexcorrectionsepitaxialgraphenePhysRevB.104.085435,Ontheoriginofcriticaltemperatureenhancementinatomicallythinsuperconductors}, we can get an effective junction length $L_{eff} \sim$ 300 nm. This value may further decrease if flux focusing is considered.
While SQUID-like patterns in single junctions are often attributed to edge-dominated supercurrents~\cite{Edgemodesuperconductivity,InducedsuperconductivityinthequantumspinHalledge,SQUIDpatterndisruptionPhysRevB.102.165407,Bi2O2SeBasedJosephsonJunction,Spatiallyresolvededgecurrentsandguidedwaveelectronicstatesingraphene,EdgesuperconductivityinmultilayerWTe2Josephsonjunction}, the extracted $I_c$($B_z$) profile [Fig.~\ref{fig:Fig_1}(c)] deviates from standard SQUID behavior: side-lobe peaks exhibit alternating amplitudes rather than monotonic decay. Such anomalies have been linked to multi-channel current distributions—e.g., three-channel cases involving crossed Andreev reflection~\cite{SuperconductingQuantumInterferencethroughTrivialEdgeStatesinInAs,Edgemodesuperconductivity,CrossedAndreevreflectioninInSbflakeJosephsonjunctions}. We address this discrepancy in later sections by analyzing the interplay between ripple-induced orbital effects and current inhomogeneity.

\begin{figure}[b]
\centering
\includegraphics[width=0.95\textwidth]{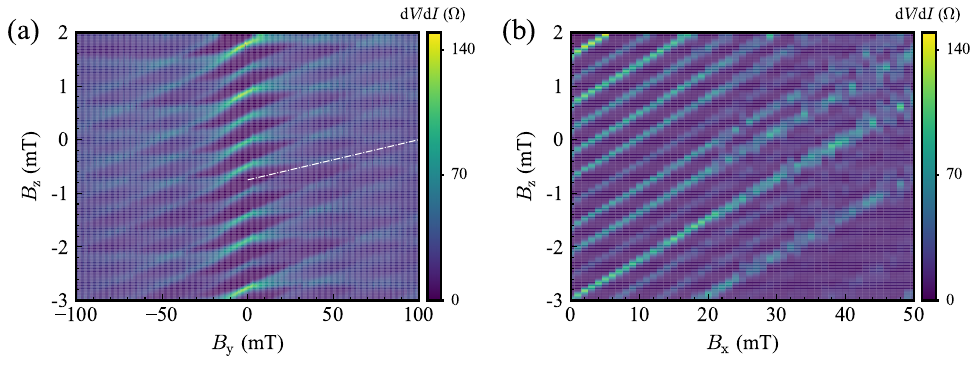}% Here is how to import EPS art
\caption{\label{fig:Fig_2} Magnetic field dependence of quantum interference patterns. The differential resistance d$V$/d$I$ map demonstrates distinct evolution characteristics under orthogonal in-plane magnetic field orientations. (a) The differential resistance d$V$/d$I$ as a function of $B_z$ and $B_y$ with $B_x$ set to zero, showing that in the presence of an in-plane magnetic field $B_y$ (transverse to the current direction), a V-shaped extension of secondary nodes is superimposed on an overall structural shift. (b) The differential resistance d$V$/d$I$ as a function of $B_z$ and $B_x$ with $B_y$ set to zero, indicating that when the in-plane magnetic field $B_x$ is parallel to the current direction, the interference pattern exhibits a simple overall shift without additional structural modifications. }
\end{figure}

\section{Transport in in-plane magnetic fields}

We next turn our focus to the response of the device to an in-plane magnetic field.
To investigate the evolution of interference patterns under in-plane magnetic fields, we employ a small AC excitation current (40 nA) to measure the differential resistance \( dV/dI \) as a function of \( B_z \) and \( B_y \) (or \( B_x \)). Consistent with the established proportionality \( dV/dI \propto 1/I_c \) in planar Josephson junctions~\cite{FinitemomentumCooperpairing,ControlledfinitemomentumpairingHgTequantumwells}, regions of reduced \( dV/dI \) correspond to enhanced critical supercurrents (\( I_c \)), while elevated \( dV/dI \) signifies \( I_c \) suppression. This methodology resolves not only the \( I_c \) modulations but also subdominant features  through its enhanced sensitivity to small current variations, circumventing the resolution limitations of direct DC measurements in low-\( I_c \) regimes.

Figure~\ref{fig:Fig_2}(a) presents the evolution of the interference pattern for junction $A$ under an in-plane magnetic field $B_y$ (perpendicular to the current direction). The application of $B_y$ induces a global shift of the interference pattern along $B_z$, manifesting as a tilted 2D $d V / d I\left(B_z, B_y\right)$ map [Fig.~\ref{fig:Fig_2}(a)]. This tilt originates from a slight sample misalignment relative to the $B_x$-$B_y$ plane, as corroborated by analogous shifts observed under $B_x$ [parallel to the current; Fig.~\ref{fig:Fig_2}(b)].
Beyond this global shift, $B_y$ drives a striking local restructuring of the interference pattern: at $B_y=0$ , nodes correspond to $d V / d I$ maxima (i.e., $I_c$ minima). As $B_y$ increases, each node bifurcates into two V-shaped branches that progressively elongate and intersect, forming a diamond lattice (Fig. 2a). Tracing $I_c$ along the diamond diagonal (white dashed line, Fig. 2a) reveals a suppression-recovery profile~\cite{supp} akin to signatures of 0-$\pi$ transitions in ferromagnet Josephson junctions~\cite{PhysRevLett.89.137007,PhysRevLett.90.167001,PhysRevLett.89.187004,PhysRevLett.96.197003,PhysRevB.73.174506,PhysRevLett.97.177003} and 2D-material Josephson junctions~\cite{ZeemanEffectInducedTransitionsDiracSemimetal,FinitemomentumCooperpairing,ControlledfinitemomentumpairingHgTequantumwells,BallisticsuperconductivityInSbquantumwells}. Crucially, this evolution is absent under $B_x$ (Fig.~\ref{fig:Fig_2}(b)), confirming its anisotropy with respect to field orientation.

Some particular types of evolution of the interference pattern in an in-plane magnetic field  has recently been predicted theoretically~\cite{CombinedZeemanandorbitaleffect} and reported experimentally in rippled graphene Josephson junctions~\cite{PlanargrapheneNbSeJosephsonjunctionsPhysRevB.103.115401}, in which the orbital effect induced by the spatial variation of rippled structures is considered to play an important role. In what follows, we give a qualitative discussion of a possible mechanism  by focusing on the orbital effect of ripples in line with the framework developed in Ref.~\cite{CombinedZeemanandorbitaleffect}.  Notably, the Zeeman effect can be safely disregarded owing to the relatively small magnetic field applied in our experiment, compared to the order of several Tesla relevant to the Zeeman-driven 0-$\pi$ transition~\cite{ZeemanEffectInducedTransitionsDiracSemimetal,BallisticsuperconductivityInSbquantumwells}.

\section{Theoretical model and discussion}

To elucidate the origin of the diamond-shaped interference pattern, we begin by considering the effect of magnetic fields on the phase of the superconducting order parameter. When a magnetic field penetrates the junction area, the Aharonov–Bohm effect induces a spatial modulation of the gauge-invariant superconducting phase difference, resulting in an oscillatory supercurrent density governed by the Josephson relation. For a junction with a sinusoidal current-phase relationship, the total Josephson current ($I_s$) is expressed as a spatial integral over the junction width $W$,
\begin{equation}\label{eq:J1}
I_s=\int_{-W / 2}^{W / 2} J_s(y) \sin \left(\varphi-\frac{2 e}{\hbar} \int_{-L / 2}^{L / 2} \mathbf{A}(x, y) d x\right) d y
\end{equation}
where $\varphi$ is the superconductor phase difference, $J_s(y)$ is the position-dependent Josephson current density with $y$ the real-space coordinate along the junction width direction, and $\mathbf{A}(x, y)$ denotes the vector potential. For a uniform current density $J_s(y)=J_0$ and a perpendicular magnetic field $B_z$, we adopt the Landau gauge $\mathbf{A}=\left(-B_z y, 0,0\right)$. This simplifies the flux integral to $\int_{-L / 2}^{L / 2} A_x d x=-B_z L y$, yielding the Fraunhofer diffraction pattern,
\begin{equation}\label{eq:J2}
I  = J_0 \int_{-W / 2}^{W / 2} \sin \left(\varphi + 2 \pi \frac{\Phi \cdot y } {\Phi_0 W}\right) d y 
=J_0 W \sin \left( \varphi \right) \frac{\sin \left(\pi \Phi / \Phi_0\right)}{\pi \Phi / \Phi_0}
\end{equation}
where $\Phi=B_z L W$ is the total magnetic flux threading the junction area, and $\Phi_0=h/2e$ is the superconducting flux quantum. This derivation recovers the canonical Fraunhofer dependence, where nodes in the critical current occur at integer multiples of $\Phi_0$.

\begin{figure}[t!]
\centering
\includegraphics[width=0.65\textwidth]{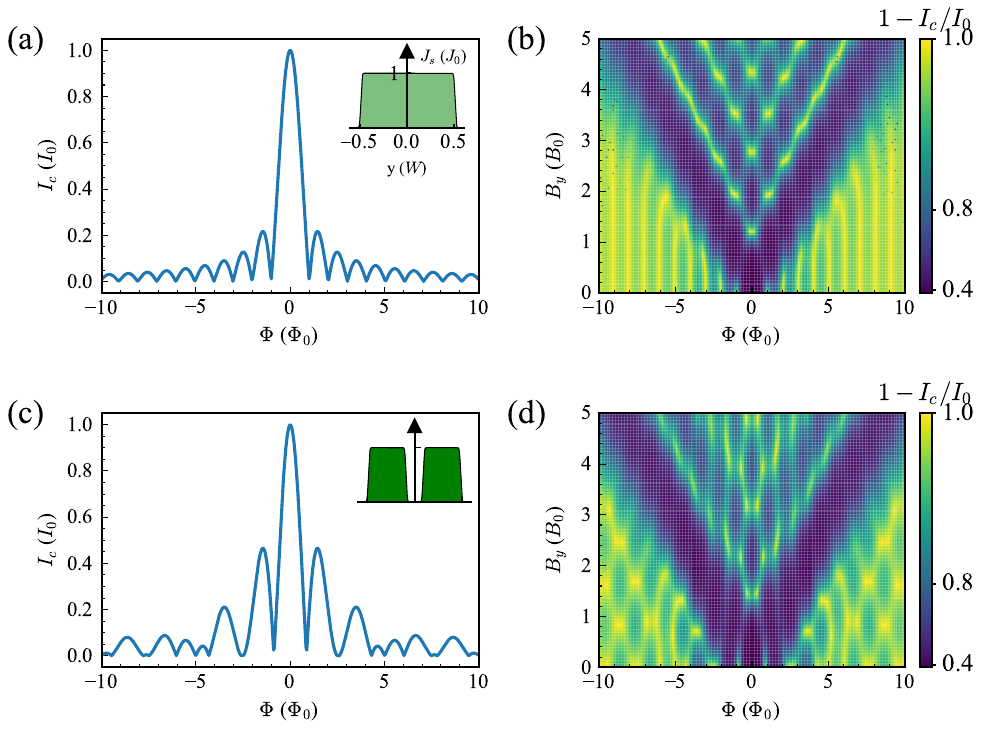}% Here is how to import EPS art
\caption{\label{fig:Fig_3} Calculated evolutions of the interference pattern as a function of in-plane magnetic field $B_y$ under specific current distribution conditions. (a) The inset illustrates a homogeneous current distribution, which exhibits a standard Fraunhofer pattern under zero in-plane magnetic field. (b) The evolution of the interference pattern under varying $B_y$ is shown for the homogeneous current distribution. The coloring indicates the critical supercurrent, with bright yellow representing zero and dark blue indicating large critical supercurrent. (c) and (d) illustrate the case of a concave-shaped current distribution. Here, $B_0$ represents the in-plane magnetic field that suppresses the critical supercurrent to zero, defined as $B_0 = \Phi_0 / 4 L \eta_0$, where $\Phi_0$ is the magnetic flux quantum, $L$ is the length scale, and $\eta_0$ is the peak-to-peak amplitude}.
\end{figure}

\begin{figure}[h]
\begin{minipage}[t]{0.48\textwidth}
% \centering
\includegraphics[width=1.0\textwidth]{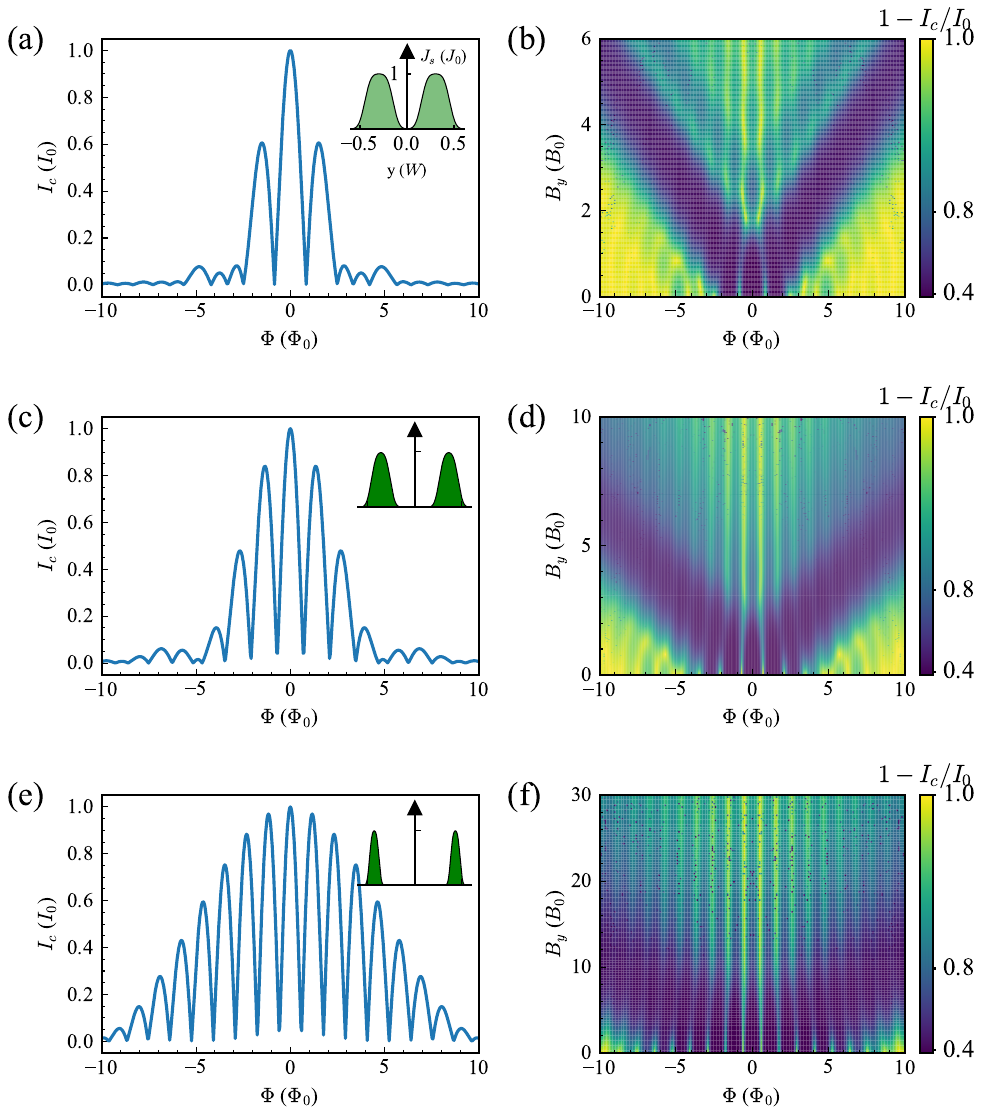}% Here is how to import EPS art
\caption{\label{fig:Fig_4} The calculated evolution of the interference pattern as a function of the in-plane magnetic field $B_y$ is shown, considering edge-dominated current density profiles. (a) The interference pattern for the specific current distribution, depicted in the inset, is displayed at zero in-plane magnetic field. (b) The evolution of the interference pattern under varying $B_y$ is demonstrated for the current distribution shown in (a). (c, d) and (e, f) present the same analysis as (a, b), but with a reduced width of the edge current density profiles.}
\end{minipage}\hspace{2pc}%
\begin{minipage}[t]{0.48\textwidth}
% \centering
\includegraphics[width=1.0\textwidth]{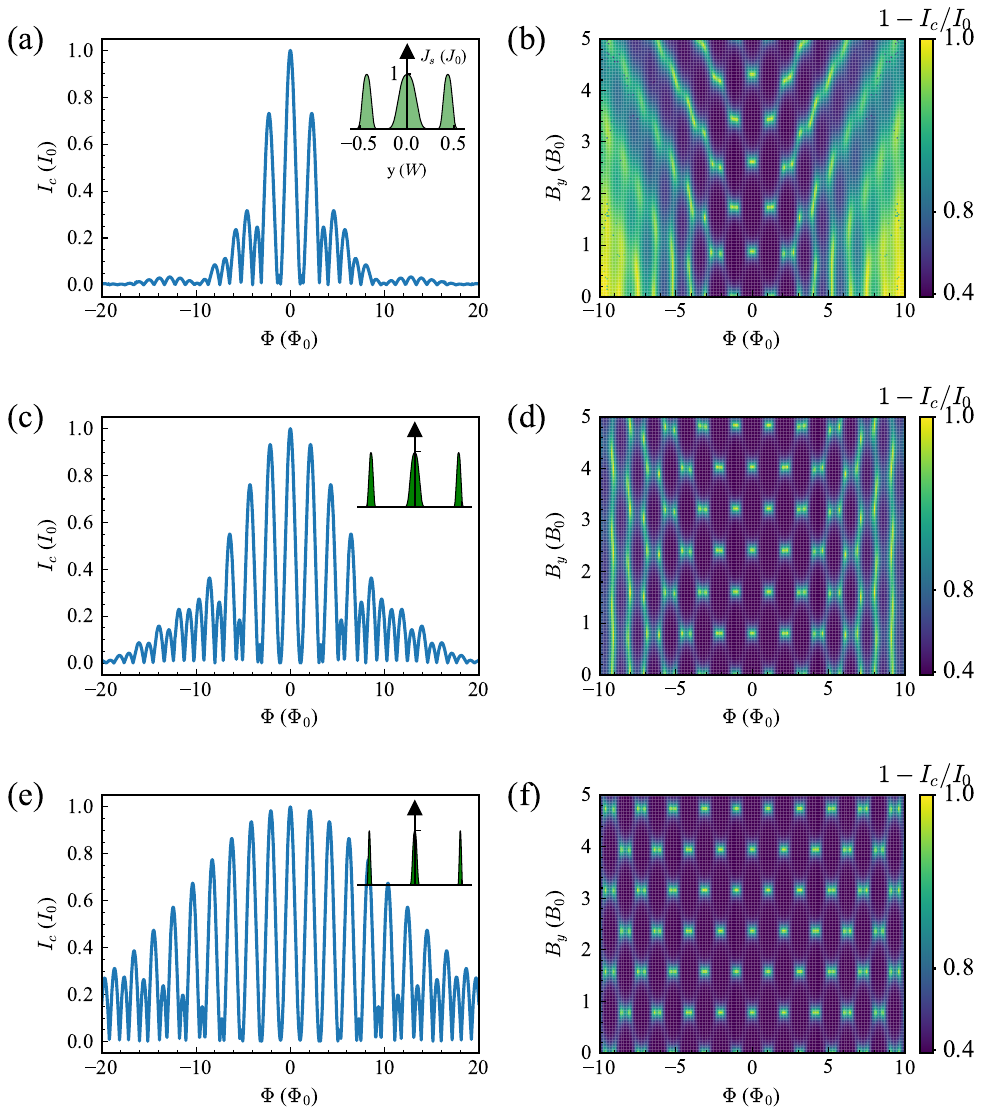}% Here is how to import EPS art
\caption{\label{fig:Fig_5}
Calculated evolution of the interference pattern as a function of $B_y$ when considering three-channel current density profiles. (a) The interference pattern for the specific current distribution sketched in the inset at zero in-plane magnetic field. (b) The evolution of the interference pattern in $B_y$ for the current distribution shown in (a). (c, d) and (e, f) The same as (a, b), but for a decreased width of the current channels.                             
}
\end{minipage} 
\end{figure}

The application of an in-plane magnetic field introduces a finite perpendicular field component due to geometric curvature arising from inherent height variations in non-ideal planar junctions. These height modulations, or rippled structures, spatially perturb the vector potential, leading to non-negligible orbital effects that significantly influence the superconducting phase coherence. As demonstrated in Ref.~\cite{CombinedZeemanandorbitaleffect}, such ripples induce oscillations in the critical supercurrent under parallel magnetic fields, a phenomenon governed by the local curvature. To model this, we parameterize the height distribution $\eta(x, y)$ of the rippled structure as,
\begin{equation}\label{eq:J3}
\eta(x, y)=\frac{\eta_0}{2} \cos \left(n \pi \frac{x}{L}\right) \cos \left(m \pi \frac{y}{W}\right)
\end{equation}
where $\eta_0$ is the peak-to-peak amplitude, and $n$, $m$ denote the number of ripples along the junction length ($x$) and width ($y$) directions, respectively. The resultant orbital effect modifies the vector potential integral as,
\begin{equation}\label{eq:J4}
\int_{-L / 2}^{L / 2} A(x, y) d x=B_y L \bar{\eta}(y)- B_z L y  
\end{equation}
where $\bar{\eta}$, the longitudinally averaged height amplitude, is given by
\begin{equation}\label{eq:J5}
\bar{\eta}(y)=\frac{1}{L} \int_{-L / 2}^{L / 2} \eta(x, y) d x.
\end{equation}
Based on Eq.~(\ref{eq:J3}), ripples are categorized into short- ($n, m \gg 1$) and long-wavelength ($n, m \sim 1$) regimes. Short-wavelength ripples produce rapidly oscillating terms in Eq. \ref{eq:J5}, which average to negligible contributions but induce orbital depairing that accelerates critical current decay~\cite{CombinedZeemanandorbitaleffect}. The threshold in-plane field $B_0$ required to suppress the critical current to zero is derived as, $B_0 = \Phi_0 / 4 L \eta_0$. For our Josephson junction (JJA) with effective length $L_{eff} \sim 300$ nm and $\eta_0 \sim 30$ nm, this yields $B_0 \sim 55$ mT, consistent with the experimental value $\sim 40$ mT. The slight discrepancy likely arises from unaccounted short-wavelength ripples—which enhance orbital depairing~\cite{CombinedZeemanandorbitaleffect}—or uncertainties in $L_{eff}$.

To investigate the origin of the diamond-shaped interference pattern, we simulate its evolution using the phenomenological model described in Eq.\ref{eq:J1}. Initially, we consider a long-wavelength ripple structure parameterized by Eq.\ref{eq:J5} with $n$ = 1, $m$ = 2, and a uniform Josephson current density. The calculated Fraunhofer-like interference under in-plane magnetic fields (Figs.\ref{fig:Fig_3} (a,b)) aligns with prior studies of ripple-induced orbital effects~\cite{CombinedZeemanandorbitaleffect}.
Subsequently, we introduce a non-uniform current distribution modeled as a two-channel step-like profile with Gaussian flat-top edges~\cite{supp} (inset, Fig.\ref{fig:Fig_3} (c)). This modification induces a striking V-shaped stretching of interference nodes [Fig.\ref{fig:Fig_3} (d)], mirroring experimental observations. The nodal evolution arises from phase interference between current channels modulated by ripple-driven orbital phases. Further narrowing the edge current profiles [Figs.\ref{fig:Fig_4} (a–d)] suppresses anti-crossing features while enhancing outward V-shaped shifts in critical current maxima—a behavior reminiscent of finite-momentum Cooper pairing mechanisms~\cite{FinitemomentumCooperpairing,ControlledfinitemomentumpairingHgTequantumwells,Josephsondiodeeffectsemimetal,4piperiodicAndreevboundstatesinaDiracsemimetalNatureMater,supp}.

Notably, for our results on JJA, the superconducting interference pattern at zero in-plane magnetic field displays a SQUID-like interference pattern but with significant discrepancies, as shown in Figs.~\ref{fig:Fig_1} (b, c). As explained above, the alternating behavior of the critical supercurrent likely indicates an underlying three-channel current distribution, similar to those reported in ~\cite{SuperconductingQuantumInterferencethroughTrivialEdgeStatesinInAs,Edgemodesuperconductivity,CrossedAndreevreflectioninInSbflakeJosephsonjunctions}. Based on this, through postulating such a three-channel profile [inset, Fig.\ref{fig:Fig_5} (a)], we achieve improved qualitative agreement with the experimental diamond-shaped pattern (Fig.\ref{fig:Fig_5} (b)). Further refining the channel width [Figs.\ref{fig:Fig_5} (c–f)] reproduces key experimental features: V-shaped nodal stretching, outward critical current shifts, and diamond-like modulation [Figs.\ref{fig:Fig_5} (e,f)], closely matching Fig.\ref{fig:Fig_2} (a).
While the three-channel model remains phenomenological, its consistency with experimental observations suggests that edge-localized current paths, likely arising from strain-induced curvature at the flake-substrate interface, may dominate transport.  Though alternative explanations cannot be entirely ruled out, key experimental signatures conflict with prominent candidates. For example,  the Little-Parks (LP) diamond effect~\cite{MurphyPhysRevB.96.094507,PhysRevApplied.16.024013KineticInductance,Babich2023LimitationsoftheCurrentPhaseRelation}, requires linear CPRs and reveals multivalued \(I_c\), which conflicts with our observations of single-valued critical currents under in-plane fields and anisotropic interference patterns~\cite{supp}. Together, these findings not only disfavor LP-like mechanisms but also underscore the dominant role of geometric curvature in governing transport anisotropy.

\section{Conclusion}

In summary, we report a distinct diamond-shaped evolution of superconducting interference patterns in NbTiN weak-link Josephson junctions under in-plane magnetic fields. This evolution is characterized by V-shaped nodal splitting and stretching, resembling suppression-recovery behaviors historically linked to Zeeman-driven 0-$\pi$ transitions. Crucially, our analysis excludes Zeeman effects—given the sub-100 mT field regime—and instead attributes the phenomenon to orbital modulation induced by nanoscale rippled structures. These geometric imperfections, arising from lateral height variations at the flake-substrate interface, makes it easier to accumulate a flux quantum within a long-wavelength ripple, and meanwhile suppresses the critical supercurrent rapidly, which can induce a suppression-recovery behavior in a relatively small magnetic field. 

By extending the phase modulation framework of Ref.~\cite{CombinedZeemanandorbitaleffect}, we incorporate spatially non-uniform current distributions into a phenomenological model. Simulations reveal that two- and three-channel step-like current profiles (edge-localized and multi-path configurations) qualitatively reproduce the experimental diamond pattern. Optimal agreement is achieved for a three-channel distribution, consistent with strain-induced edge currents in the junction’s non-ideal geometry. Our findings serve as an extended understanding of rich orbital interference effects in Josephson junctions, revealing how rippled structures and non-uniform current distributions reshape phase-coherent transport. This cautions against misinterpreting similar patterns as Zeeman-driven phenomena in planar junctions, with implications for designing and interpreting superconducting devices under in-plane fields.

\section*{Data availability statement}
The data that support the findings of this study are available upon reasonable request from the authors.

\section*{Acknowledgement}
This work was supported by the National Key Research and Development Program of China (2022YFA1403400), by the NSF China (92365207, 12074417, 92065203, and 11774405), by the Strategic Priority Research Program B of Chinese Academy of Sciences (XDB28000000 and XDB33000000), by the Synergetic Extreme Condition User Facility sponsored by the National Development and Reform Commission, and by the Innovation Program for Quantum Science and Technology (2021ZD0302600).

\nocite{*}

\section*{References}
%%%=============================================
%%%=============================================
% \bibliography{main}% Produces the bibliography via BibTeX.
% \input{main.bbl}

\providecommand{\noopsort}[1]{}\providecommand{\singleletter}[1]{#1}%
\providecommand{\newblock}{}

%%%============================================
\end{document}

%======================== head
\preprint{APS/123-QED}
\title{Supplemental Material for: "Diamond-shaped superconducting interference patterns in NbTiN weak-link Josephson junctions under in-plane magnetic fields"}% Force line breaks with \\

\author{Kui Zhao}
    % \thanks{zhaokui@baqis.ac.cn}
        \altaffiliation{These authors contributed equally to this work.}
	\affiliation{\addrA}
    \affiliation{\addrD}
\author{Jianfei Xiao}
    % \thanks{}
        \altaffiliation{These authors contributed equally to this work.}
	\affiliation{\addrA}
        \affiliation{\addrB}
\author{Huaiyuan Liu}
    % \thanks{}
        \altaffiliation{These authors contributed equally to this work.}
	\affiliation{\addrA}
\author{Linfeng Tu}
    % \thanks{}
	\affiliation{\addrA}
        \affiliation{\addrB}
\author{Yiwen Ma}
    % \thanks{}
	\affiliation{\addrA}
        \affiliation{\addrB}
\author{Jiangbo He}
    % \thanks{}
	\affiliation{\addrA}
\author{Mingli Liu}
    % \thanks{}
	\affiliation{\addrA}
        \affiliation{\addrB}
\author{Ruiyang Jiang}
    % \thanks{}
	\affiliation{\addrA}
        \affiliation{\addrB}
\author{Zhongmou Jia}
    % \thanks{}
	\affiliation{\addrA}
        \affiliation{\addrB}
\author{Shang Zhu}
    % \thanks{}
	\affiliation{\addrA}
        \affiliation{\addrB}
\author{Yunteng Shi}
    % \thanks{}
	\affiliation{\addrA}
        \affiliation{\addrB}
\author{Zhaozheng Lyu}
    % \thanks{}
	\affiliation{\addrA}
        \affiliation{\addrC}
\author{Jie Shen}
    % \thanks{}
	\affiliation{\addrA}
\author{Guangtong Liu}
    % \thanks{}
	\affiliation{\addrA}
        \affiliation{\addrB}
        \affiliation{\addrC}
\author{Li Lu}
    \thanks{lilu@iphy.ac.cn}
	\affiliation{\addrA}
        \affiliation{\addrB}
        \affiliation{\addrC}
\author{Fanming Qu}
    \thanks{fanmingqu@iphy.ac.cn}
	\affiliation{\addrA}
        \affiliation{\addrB}
        \affiliation{\addrC}
% \date{\today}
%=======================Abstract=====================
% \begin{abstract}

% \end{abstract}
\maketitle

\tableofcontents

%\tableofcontents
%========================Introduction====================

\subsection{Gaussian flat-top profile}

Here we adopt a Gaussian flat-top functions to portray the current density distributions,  in order to adapt to the realistic situation as much as possible. The density profile is parameterized by the relative width and weights of each constituent component as follows,
\begin{equation}\label{eq:Gaussian_01}
\begin{aligned}
J_s(y)= J_{\mathrm{bulk}}+\sum_{i}\frac{J^i_{\mathrm{step}}-J_{\mathrm{bulk}}}{2}\left[\operatorname{Erf}\left(\frac{y - ( y^i_0-\frac{1}{2} y^i_{\mathrm{width}} )}{\sqrt{2} \sigma_i}\right)\right. 
 \left.-\operatorname{Erf}\left(\frac{y - (y^i_0 + \frac{1}{2} y^i_{\mathrm{width}} )}{\sqrt{2} \sigma_i}\right)\right],
\end{aligned}
\end{equation}
%In Eq.~(\ref{eq:Gaussian_01}) 
where,Erf is standard Gauss error function, $y^i_{\mathrm{width}}$ parameterize the width of i-th part density profile with its central points located at $y^i_0$ and corresponding supercurrent density $J^i_{\mathrm{step}}$ at this point. The bulk current density $J_{\mathrm{bulk}}$ represents the baseline contribution outside the channels. The smoothing parameter$\sigma_i$ controls the edge sharpness of each channel (Fig.~\ref{fig:FigS_00}), enabling tunable transitions between step-like and graded profiles. This formula provides versatile control over current distribution morphology—from localized edge channels to broad bulk-dominated flows—allowing systematic exploration of interference pattern evolution under varying in-plane magnetic fields.

\subsection{Evolution of the interference pattern with three-channel current density profiles}

Beyond the cases presented in the main text, we systematically investigate the interplay between current distribution morphology and interference pattern evolution under in-plane magnetic fields (Figs.~\ref{fig:FigS_00}\textcolor{blue}{-5}). Each current density profile generates a unique evolutionary signature when associated with ripple-induced orbital effects, effectively serving as a fingerprint for the underlying transport configuration.

For three-channel profiles with uniform width (Figs.~\ref{fig:FigS_00}), the interference pattern evolves symmetrically with increasing $B_y$. Introducing lateral offsets to the central channel (Fig.~\ref{fig:FigS_01}) induces a characteristic tilt in the nodal structure, marked by the dashed red guidelines. Further investigations into variations in central channel width and relative weights of current distributions reveal that the spatial distribution of the central channel significantly influences the evolution of the interference pattern. As shown in Fig.~\ref{fig:FigS_02}\textcolor{blue}{-5}, narrower lobes enhance nodal splitting effects, while asymmetric weighting induces asymmetric splitting of critical current maxima. These results establish a quantitative correspondence between current distribution parameters (width, position, weighting) and interference pattern geometry. The sensitivity of nodal evolution to central lobe modifications underscores the critical role of edge-localized currents in ripple-mediated phase modulation.

 \begin{figure}[H]
\centering
\includegraphics[width=0.4\textwidth]{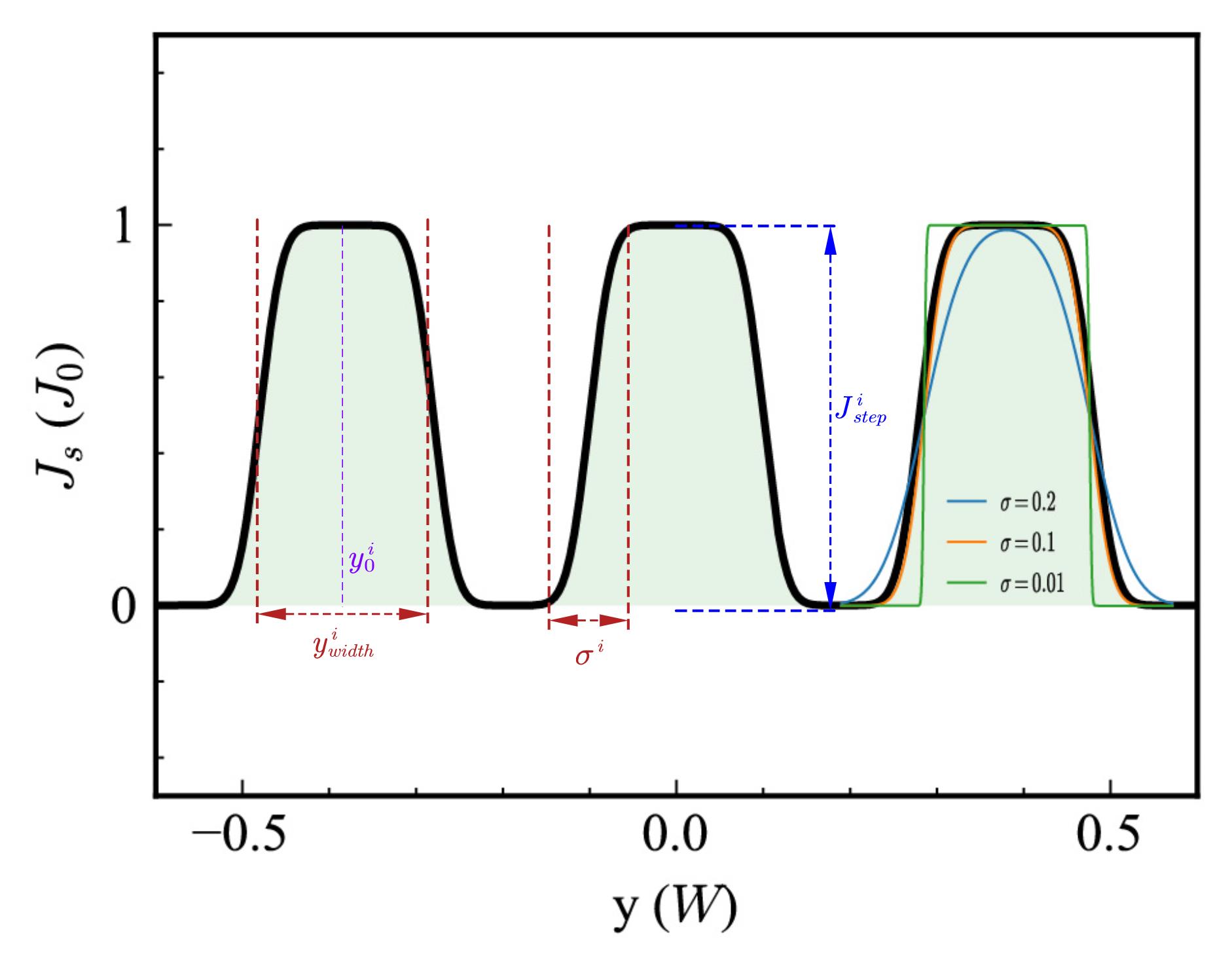}% Here is how to import EPS art
\caption{\label{fig:FigS_00}
A schematic of the current density distribution $J_s$ along the $y$ axis, dictated by Gaussian flat-top functions. }
\end{figure}

\begin{figure}[H]
	\centering
	\begin{minipage}{0.48\textwidth}
		\centering
		\includegraphics[width=1.0\linewidth]{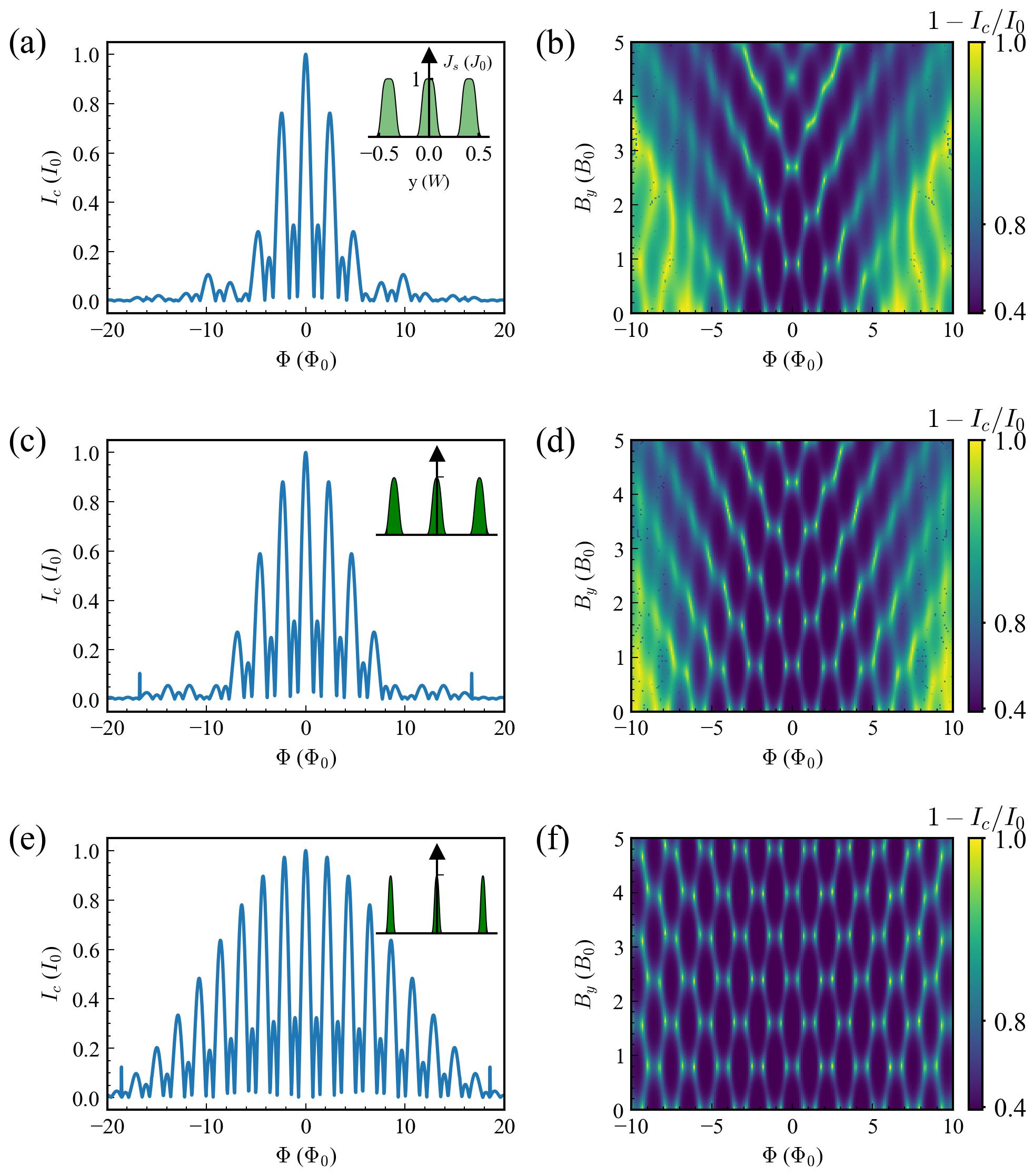}
		\caption{Calculated evolution of the interference pattern as a function of $B_y$ when considering three-channel current density profiles.}
		\label{fig:FigS_00}%文中引用该图片代号
	\end{minipage}
        \hfill
	\begin{minipage}{0.48\textwidth}
		\centering
		\includegraphics[width=1.0\linewidth]{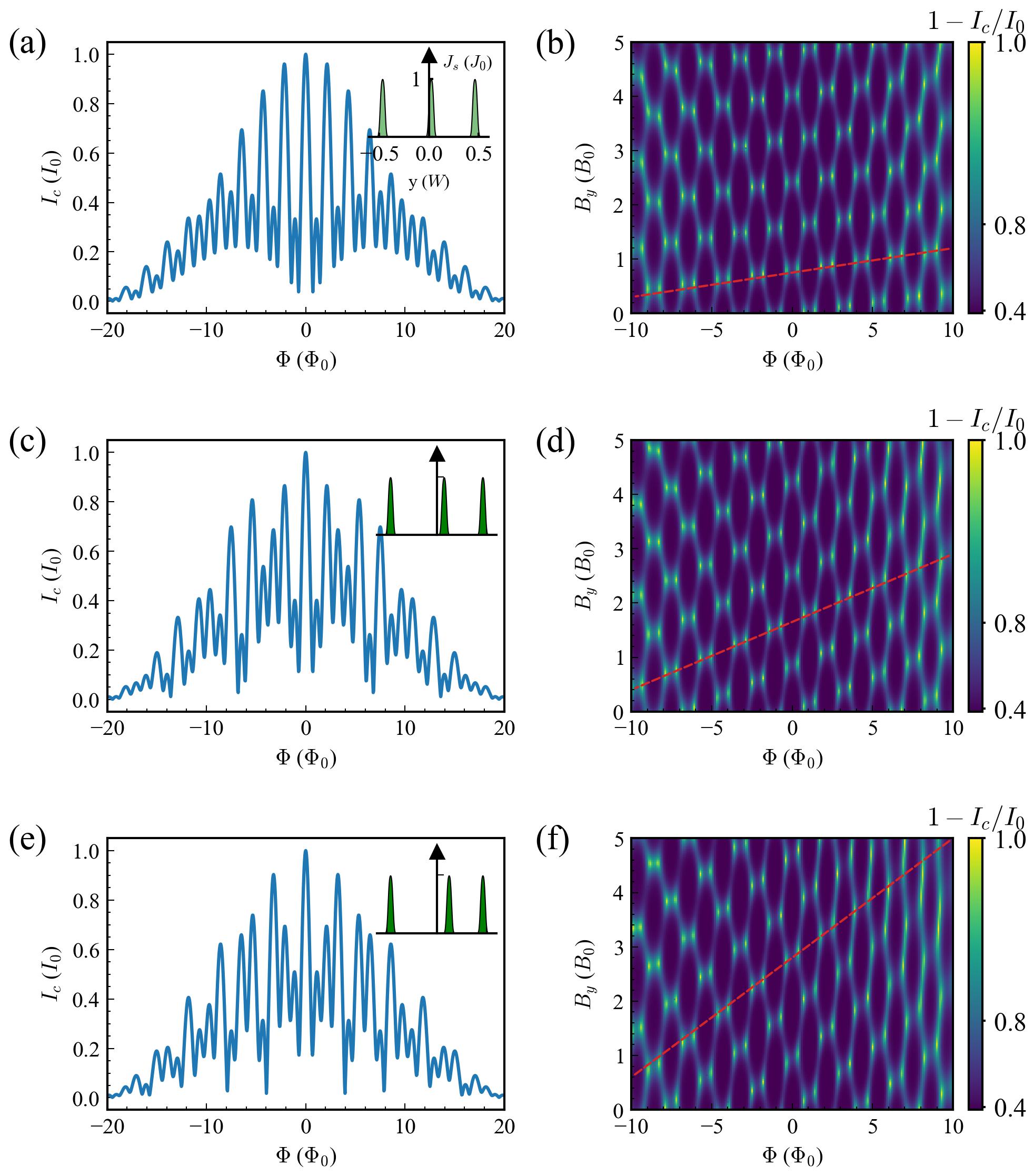}
		\caption{Calculated evolution of the interference pattern as a function of $B_y$ when considering three-channel current density profiles.The red dash line shows a guideline for the overall tilting of the evolution pattern as the position of the central current step deviates from the center of the junction.}
		\label{fig:FigS_01}%文中引用该图片代号
	\end{minipage}
	%\qquad
	%让图片换行，
\end{figure}

\begin{figure}[H]
	\centering
	\begin{minipage}{0.48\textwidth}
		\centering
		\includegraphics[width=1.0\linewidth]{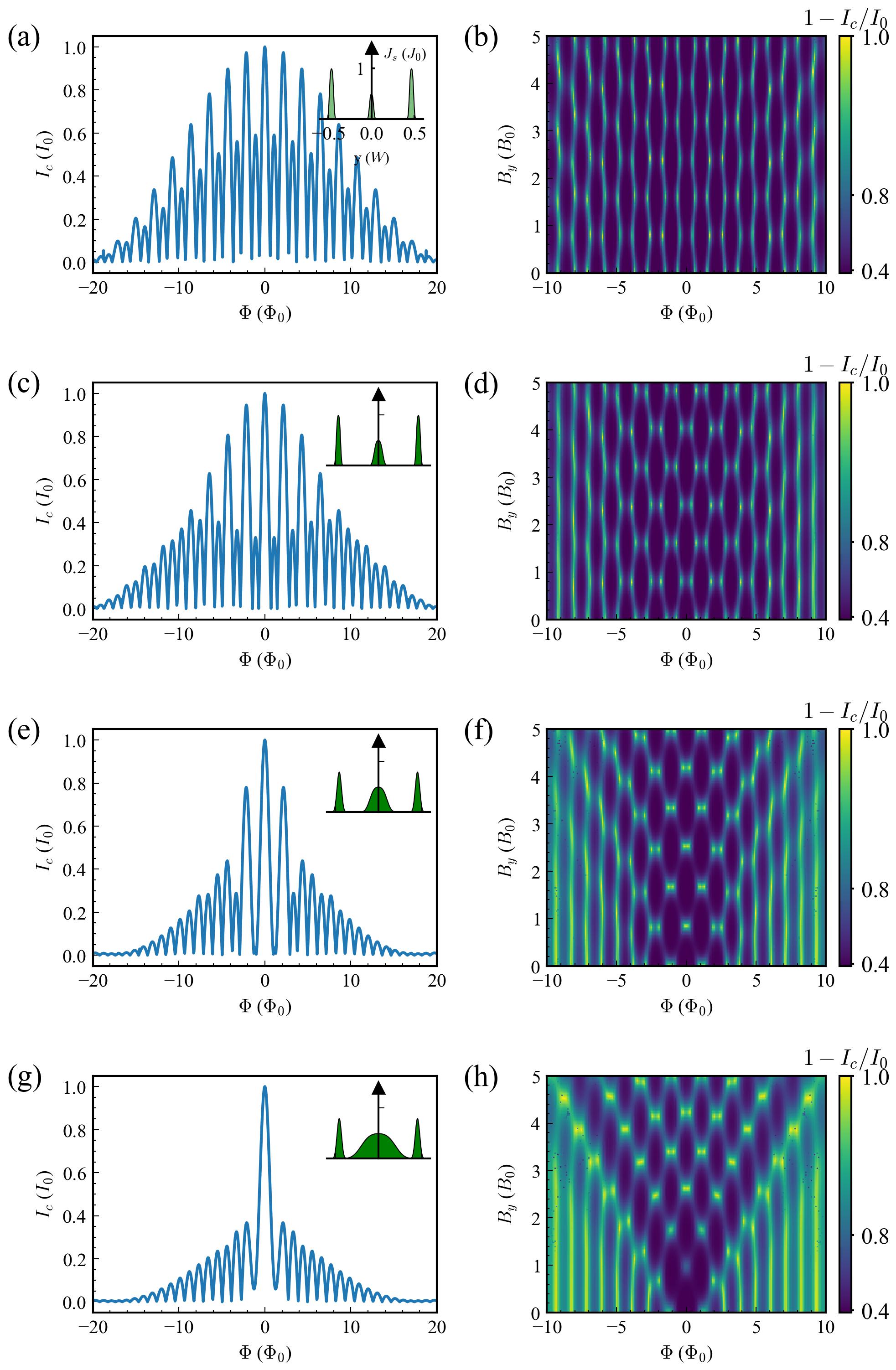}
		\caption{Calculated evolution of the interference pattern as a function of $B_y$ when considering three-channel current density profiles.}
		\label{fig:FigS_02}%文中引用该图片代号
	\end{minipage}
        \hfill
	\begin{minipage}{0.48\textwidth}
		\centering
		\includegraphics[width=1.0\linewidth]{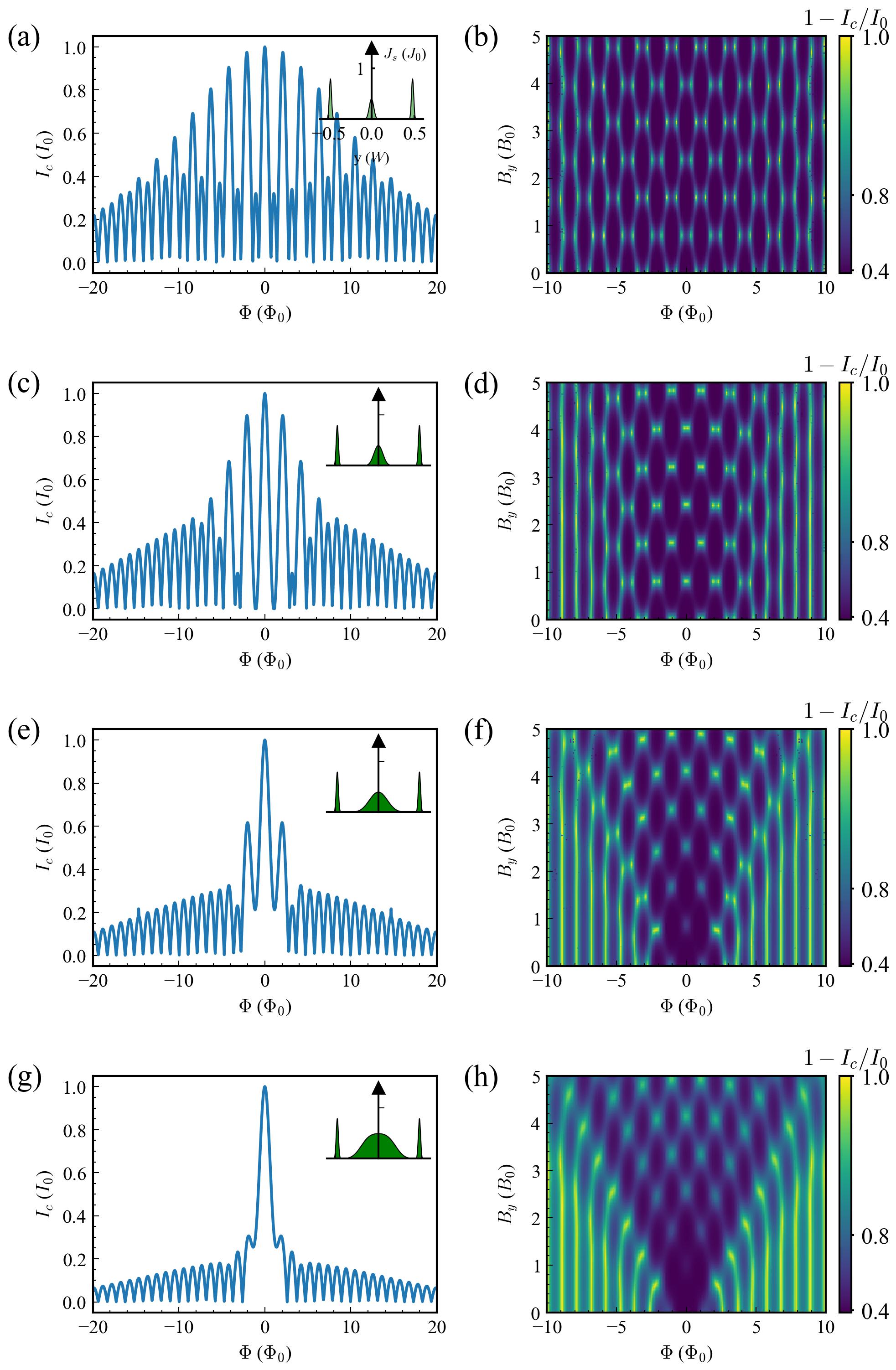}
		\caption{The calculated evolution of the interference pattern in response to variations in $B_y$, considering three-channel current density profiles.}
		\label{fig:FigS_03}%文中引用该图片代号
	\end{minipage}
	%\qquad
	%让图片换行，
\end{figure}

\subsection{Evolution of the interference pattern with edge dominated current density profiles}

\begin{figure}[H]
	\centering
	\begin{minipage}{0.43\textwidth}
		\centering
		\includegraphics[width=1.0\linewidth]{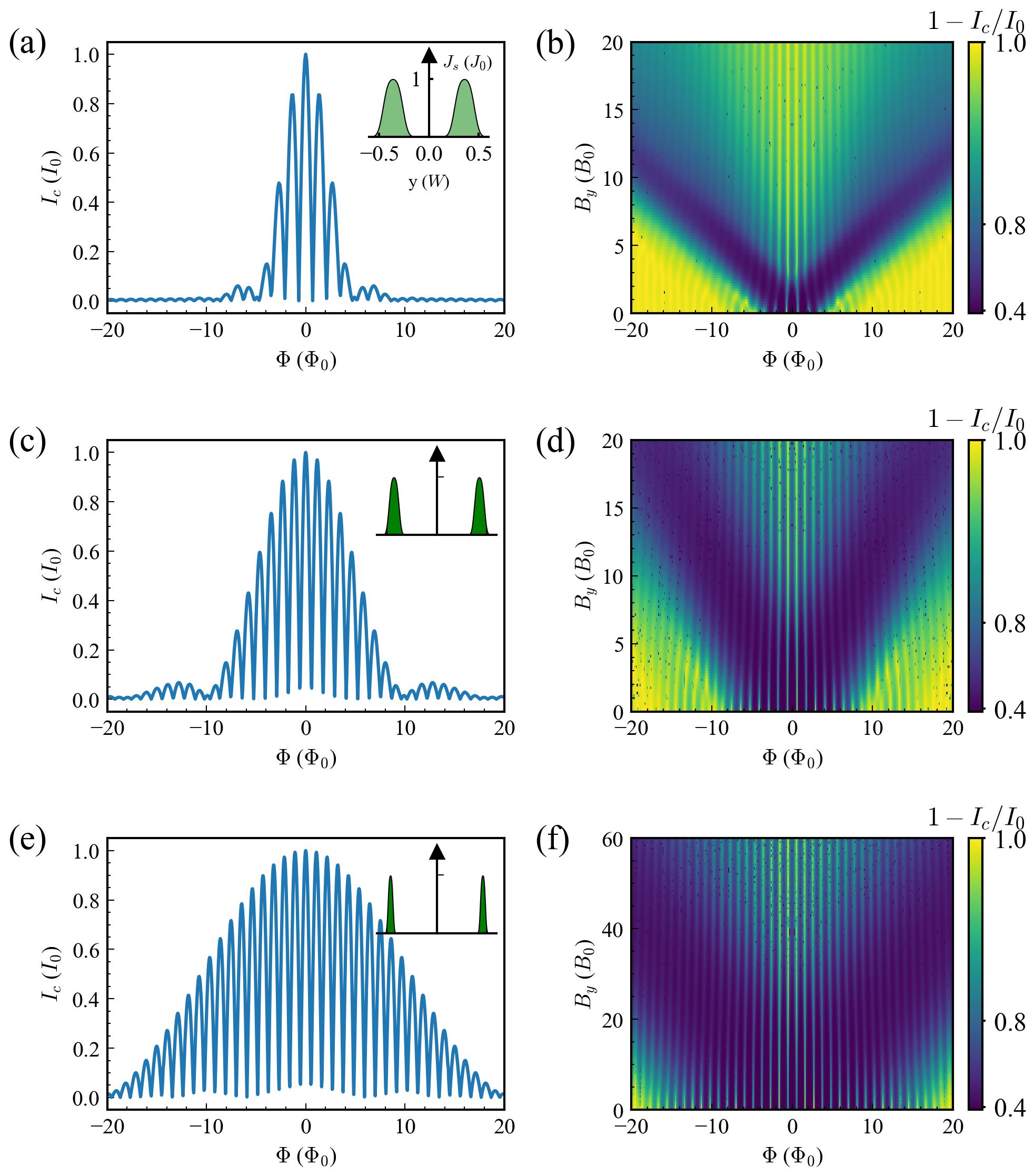}
		\caption{The calculated evolution of the interference pattern as a function of $B_y$, with edge-dominated current density profiles, shows that the patterns in (a,b), (c,d), and (e,f) evolve as the edge current density width decreases.}
        \label{fig:figS_GaussFlattopStep2_001}%文中引用该图片代号
	\end{minipage}
        \hfill
	\begin{minipage}{0.54\textwidth}
		\centering
		\includegraphics[width=1.0\linewidth]{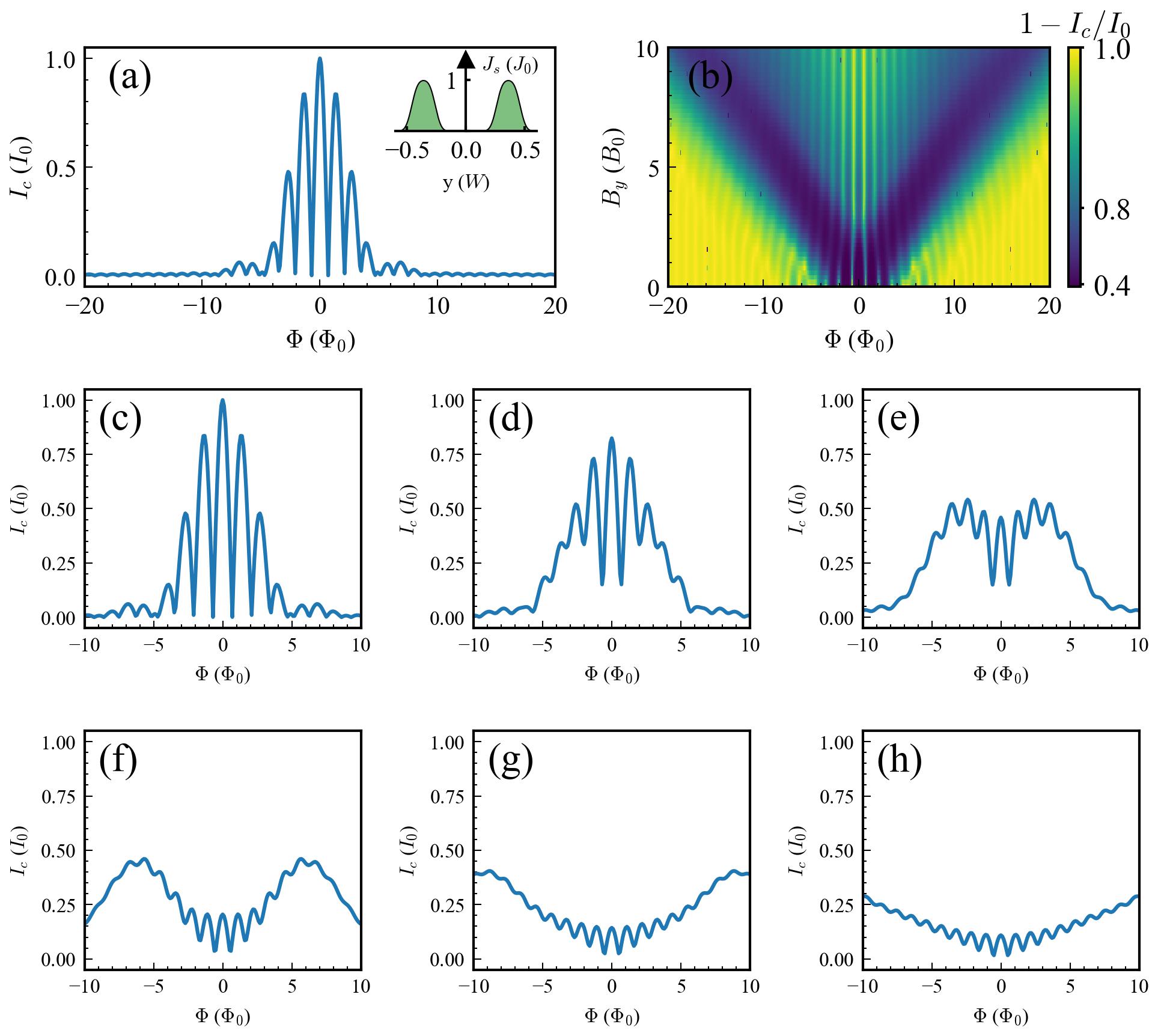}
		\caption{The calculated evolution of the interference pattern as a function of $B_y$, consistent with Figs.~\ref{fig:figS_GaussFlattopStep2_001}(a) and (b). Critical current $I_c$ versus $B_z$ for varying $B_y$ values, as shown in (c)-(h), reveals that the intensity of the supercurrent shifts from the center to the sides with increasing $B_y$.}
        \label{fig:figS_GaussFlattopStep2_001-1}%文中引用该图片代号
	\end{minipage}
	%\qquad
	%让图片换行，
\end{figure}

In Fig.~\ref{fig:figS_GaussFlattopStep2_001}, we present the calculated evolution of the interference pattern as a function of $B_y$ for edge-dominated current density profiles. Cross-sectional slices at fixed $B_y$ values (Fig.~\ref{fig:figS_GaussFlattopStep2_001-1}) reveal a progressive transfer of supercurrent intensity from the central lobe to side branches with increasing field strength. This redistribution of critical current mirrors experimental signatures  attributed to finite-momentum Cooper pairing~\cite{FinitemomentumCooperpairing,ControlledfinitemomentumpairingHgTequantumwells,Josephsondiodeeffectsemimetal,4piperiodicAndreevboundstatesinaDiracsemimetalNatureMater}.
 Here our results suggest that the orbital effect combined with general edge dominated current density distributions can also cause similar evolution of the interference pattern related to finite momentum Cooper pairing. However, it's rather easy to distinguish between these two behaviors, because the former requires extreme conditions comprising strong magnetic field, inhomogeneous current distribution and relatively large span spatial geometric structure.

 Our analysis demonstrates that orbital effects, when coupled with edge-localized current distributions, can produce interference patterns similar to those arising from finite-momentum pairing. However, the two mechanisms are readily distinguishable, the former requires extreme conditions—strong magnetic fields, highly inhomogeneous current distributions, and large spatial geometries—whereas orbital effects dominate in moderate fields and nanoscale junctions. This distinction underscores the necessity of carefully characterizing current profiles and geometric imperfections when interpreting interference phenomena in planar Josephson systems.

\subsection{ Temperature dependence of critical supercurrent}

\begin{figure}[H]
\centering
\includegraphics[width=0.75\textwidth]{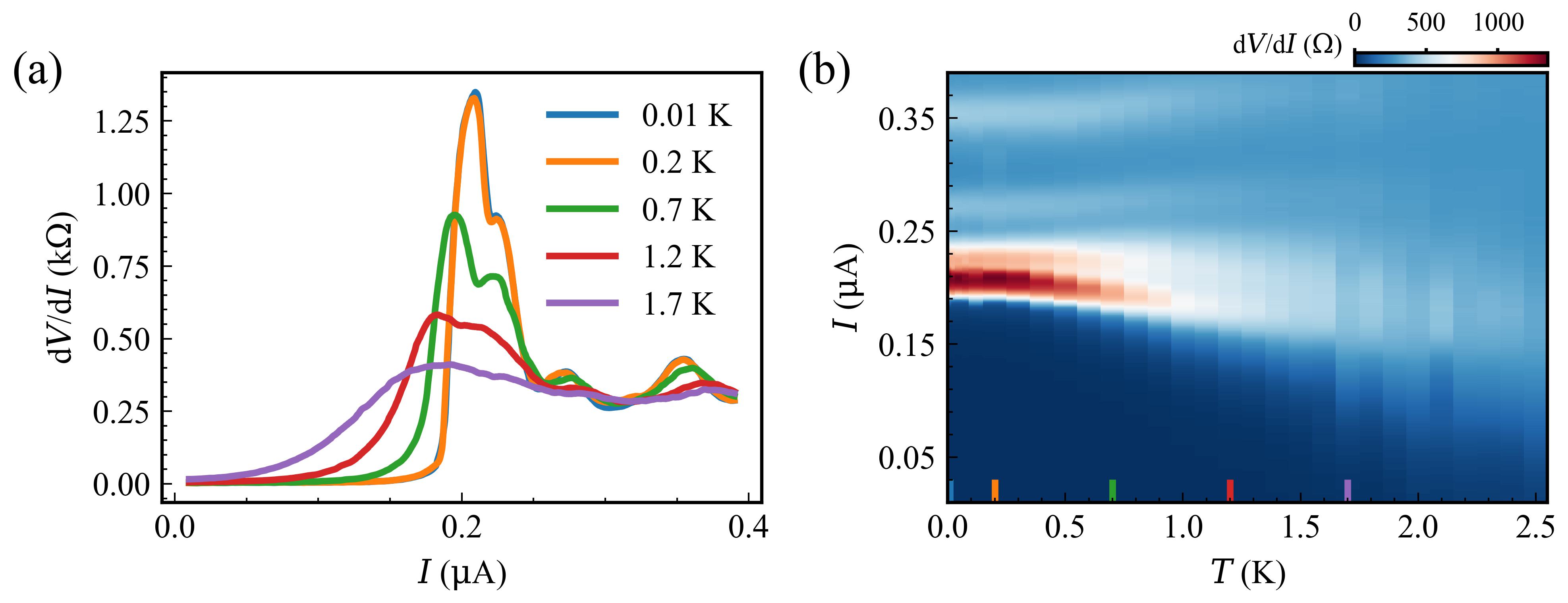}% Here is how to import EPS art
\caption{\label{fig:FigS_T_Ic}
The temperature dependence of the critical Supercurrent of JJA. (a) Differential voltage (d$V$/d$I$) versus current ($I$) measurements at different temperatures, as indicated by the colored ticks in (b). (b) Color map of d$V$/d$I$ versus $I$ and temperature ($T$).}
\end{figure}

\subsection{Superconducting interference spectra at different in-plane magnetic field $\mathit{B_y}$}

\begin{figure}[H]
\centering
\includegraphics[width=0.8\textwidth]{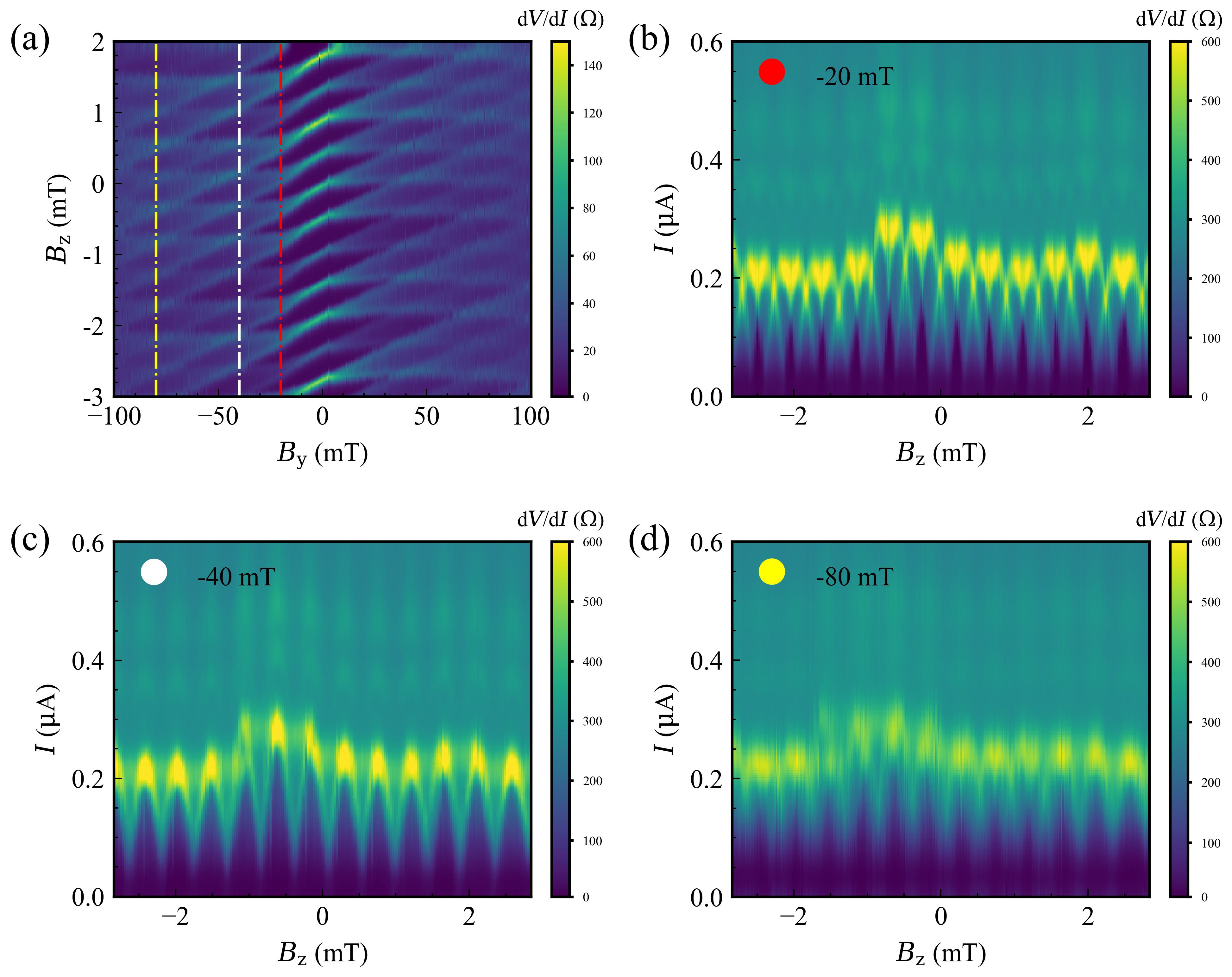}% Here is how to import EPS art
\caption{\label{fig:FigS_Part1}
The superconducting interference spectra in different in-plane magnetic fields $B_y$. (a) Evolution of the interference pattern in an in-plane magnetic field $B_y$, the same as Fig. 2(a) in the main text. (b–d) Differential resistance $dV/dI$ as a function of perpendicular magnetic field $B_z$ and bias current $I$ at $B_y$ = -20, -40, -80 mT, respectively, as indicated by the dot-dashed lines in (a).}
\end{figure}

Superconducting interference spectra at specific in-plane magnetic fields $B_y$ are shown in Fig.~\ref{fig:FigS_Part1}. The interference spectra exhibit a notable evolutionary behavior. A distinct suppression-and-revival tendency of the critical supercurrent can be observed.

\begin{figure}[H]
\centering
\includegraphics[width=0.8\textwidth]{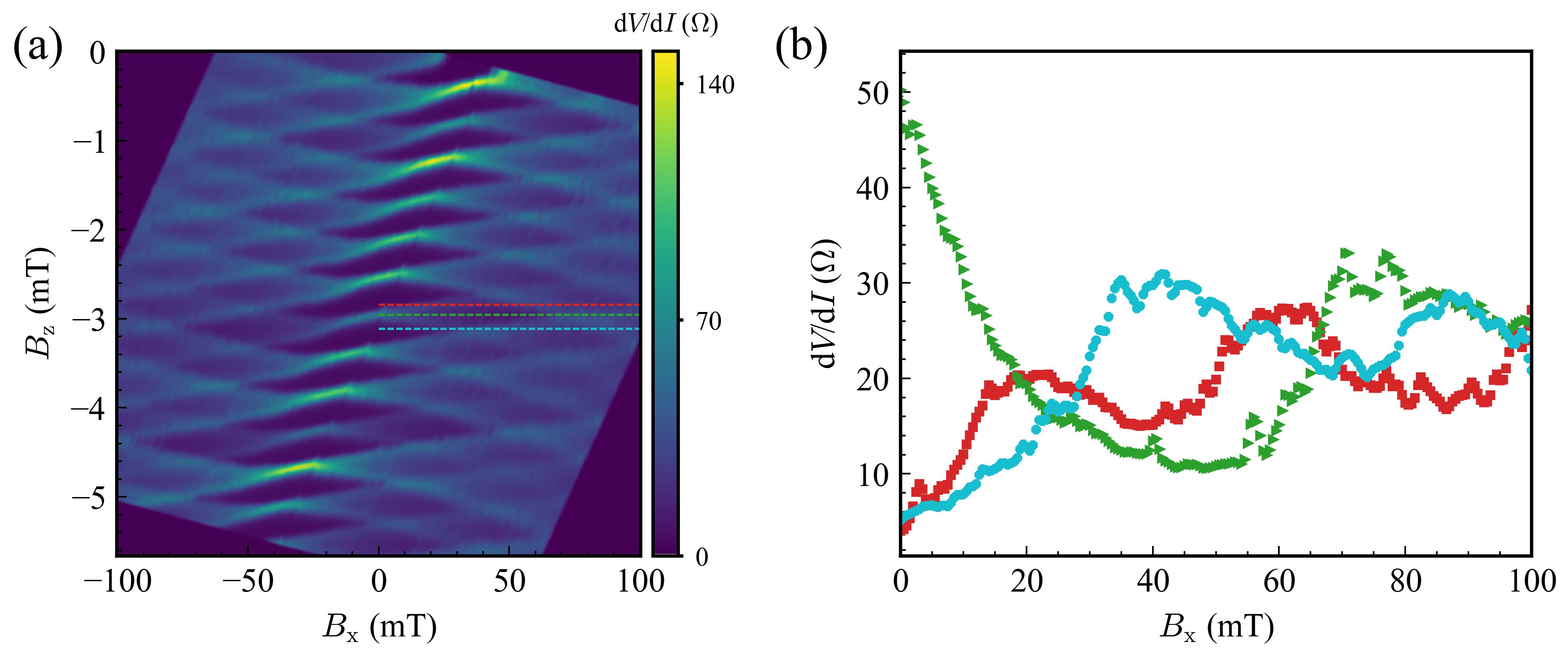}% Here is how to import EPS art
\caption{\label{fig:Fig_Diamond1-rotate}
Superconducting interference spectra under varying in-plane magnetic fields \( B_y \). (a) Differential resistance (\( dV/dI \)) map as a function of \( B_x \) and \( B_y \), corrected for a slight sample misalignment relative to the \( B_x-B_y \) plane through coordinate rotation. (b) Extracted \( dV/dI \) curves (colored lines in (a)) at distinct regions of the diamond-shaped pattern: cyan (vertex), red (mid-edge), and green (intersection), demonstrating distinct modulation behaviors under parallel field tuning. Since \( dV/dI \) is inversely proportional to the critical current \( I_c \) (\( dV/dI \propto 1/I_c \)), the experimental data exhibit qualitative agreement with simulated \( I_c \) interference patterns, capturing the suppression and recovery features across the diamond structure.
}
\end{figure}

The superconducting interference spectra under varying in-plane magnetic fields \( B_y \) reveal distinct modulation behaviors in the differential resistance (\( dV/dI \)) maps. Figure~\ref{fig:Fig_Diamond1-rotate}(a) presents the \( dV/dI \) as a function of \( B_x \) and \( B_y \), corrected via coordinate rotation to account for a slight sample misalignment relative to the \( B_x-B_y \) plane. This adjustment ensures the observed diamond-shaped interference pattern accurately reflects the intrinsic response to in-plane fields. Fig.~\ref{fig:Fig_Diamond1-rotate}(b), extracted \( dV/dI \) curves, corresponding to specific regions of the diamond structure (cyan: vertex, red: mid-edge, green: intersection), demonstrate how parallel field tuning modulates the \( dV/dI \) suppression and recovery. These features align qualitatively with simulated \( I_c \) interference patterns, leveraging the inverse proportionality \( dV/dI \propto 1/I_c \)~\cite{FinitemomentumCooperpairing,ControlledfinitemomentumpairingHgTequantumwells}. The diamond-shaped suppression pattern, governed by orbital effects from interfacial height variations, underscores the dominant role of nanoscale topography in shaping anisotropic superconducting transport under in-plane magnetic fields.

\subsection{Experimental Distinction from Little-Parks Vorticity Diamonds}

% Asymmetric nanowire SQUID: Linear current-phase relation, stochastic switching, and symmetries~\cite{MurphyPhysRevB.96.094507}.

% Impact of Kinetic Inductance on the Critical-Current Oscillations of Nanobridge SQUIDs~\cite{PhysRevApplied.16.024013KineticInductance}.

% Limitations of the Current–Phase Relation Measurements by an Asymmetric dc - SQUID~\cite{Babich2023LimitationsoftheCurrentPhaseRelation}.

% Current–Phase Relation of a WTe${_2}$ Josephson Junction~\cite{Endres2023CurrentPhaseRelationWTe}.

The Little-Parks (LP) diamond effect~\cite{MurphyPhysRevB.96.094507,PhysRevApplied.16.024013KineticInductance,Babich2023LimitationsoftheCurrentPhaseRelation,Endres2023CurrentPhaseRelationWTe}, reported in SQUIDs with superconducting weak links, manifests as multivalued critical currents and stochastic switching behavior governed by vorticity transitions. This phenomenon is captured by the vorticity diamond model~\cite{MurphyPhysRevB.96.094507,PhysRevApplied.16.024013KineticInductance}, which assumes linear current-phase relations (CPRs) in the weak links: \( I_j = (I_{cj}/\varphi_{cj})\varphi_j \), where \( I_{cj} \) and \( \varphi_{cj} \) denote the critical current and critical phase, respectively. The total critical current \( I_c(B, n_v) \), determined by the phase quantization condition \( \varphi_1 - \varphi_2 + 2\pi B/\Delta B = 2\pi n_v \), forms diamond-shaped patterns for each vorticity state \( n_v \). These diamonds arise from the linear CPRs of the weak links and overlap when the total critical phase \( \varphi_{c,\text{tot}} = \varphi_{c1} + \varphi_{c2} > \pi \), leading to multivalued critical currents in overlapping field regions. Each diamond branch reflects the linear CPR of one weak link~\cite{MurphyPhysRevB.96.094507,PhysRevApplied.16.024013KineticInductance}. 

Our experimental observations differ fundamentally from this framework. The Little-Parks diamond mechanism requires a perpendicular magnetic field (\( B_z \)) to modulate the phase winding around the SQUID loop, relying on linear CPRs to generate overlapping diamonds. However, our device exhibits typical SQUID-like interference patterns under \( B_z \), inconsistent with the predicted linear CPR-induced diamond structure (Fig. 1a in the main text). This discrepancy suggests a conventional sinusoidal CPR in our weak links rather than a linear one. Crucially, the interference diamonds we observe are modulated by in-plane magnetic fields (\( B_{||} \)) and exhibit pronounced directional anisotropy (Fig. 2), features incompatible with the \( B_z \)-driven LP paradigm. Furthermore, the critical current in our measurements remains single-valued at fixed \( B_{||} \), lacking the multivalued behavior expected from overlapping vorticity diamonds. These deviations confirm that the vorticity diamond mechanism, which hinges on linear CPRs and perpendicular-field-induced phase winding, cannot explain our results. The observed phenomena instead point to a conventional Josephson junction-like response with sinusoidal CPRs and in-plane flux-driven orbital effects unique to the device geometry.

\subsection{Current - voltage characteristic and resonances above \( I_c \)}

% Fractional AC Josephson effect in a topological insulator proximitized by a self-formed superconductor~\cite{PhysRevB.110.064511}.

% Transparent Semiconductor-Superconductor Interface and Induced Gap in an Epitaxial Heterostructure Josephson Junction~\cite{PhysRevApplied.7.034029}.

\begin{figure}[H]
\centering
\includegraphics[width=0.95\textwidth]{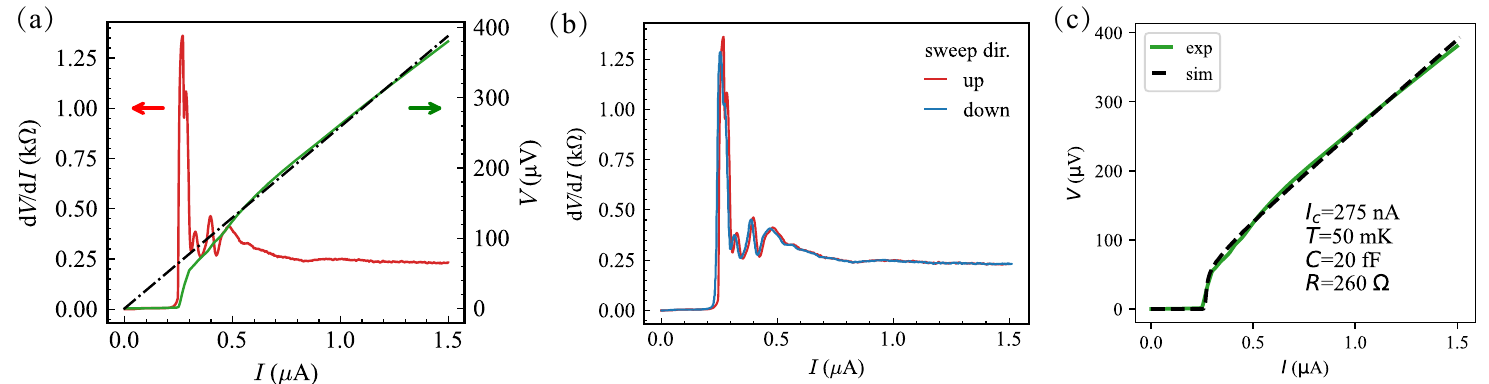}% Here is how to import EPS art
\caption{\label{fig:FigS_IV_hysteresis}
(a) (Left) Differential resistance (d$V$/d$I$) versus current bias; (Right) DC voltage derived through numerical integration of d$V$/d$I$ data. Linear fits at high bias are shown as dashed lines. The value of the excess current is obtained from the 
$V$=0 intercept of the linear fit.(b) Current-voltage characteristics during upward (red line) and downward (blue line) DC-bias sweeps, exhibiting minimal hysteresis between scan directions. (c) Good agreement between the RSJ model and the V-I curve in panel (a) is observed.}
\end{figure}

Figure~\ref{fig:FigS_IV_hysteresis}(a) displays the differential resistance (left axis) and DC voltage (right axis) as functions of applied current. At higher currents, the differential resistance saturates at the normal-state resistance, with the critical current \( I_c = 275\ \text{nA} \) and normal-state resistance \( R_n \approx 260\ \Omega \) yielding a characteristic \( I_cR_n \) product of \( 71.5\ \mu\text{eV} \). The excess current \( I_e \) is determined by linearly extrapolating the high-voltage regime  to zero bias (black dashed line). However, since the experimental results do not fully meet the strict voltage requirement ( \( V \gg \Delta/e \)~\cite{PhysRevApplied.7.034029,wiedenmann20164pi}), the obtained \( I_e \) value is considered only as a reference. The extracted excess current \( I_e \) approaches zero, which is indicative of a low junction transparency, based on the Blonder-Tinkham-Klapwijk (BTK) theory~\cite{BTKPhysRevB.99.245302,BTKPhysRevB.38.8707,BTKPhysRevB.27.6739}.

The differential resistance peaks observed above the critical current \( I_c \) (in main text Fig. 1(a,b)) exhibit behavior distinct from conventional multiple Andreev reflections (MAR). Similar features were reported in topological Josephson junctions~\cite{PhysRevB.110.064511}. As shown in Fig.~\ref{fig:FigS_IV_hysteresis}, the current-voltage hysteresis remains minimal, while additional differential resistance peaks emerge at \( |I| > I_c \). Notably, these peaks occur at fixed voltage positions despite their field-dependent current bias shifts (see Main Text Fig. 1(a,b)) and show no temperature shift below \( T_c \) (Fig.~\ref{fig:FigS_T_Ic}), excluding both Fiske steps and coherence-related Andreev reflection mechanisms. Therefore, we consider that the resonant behavior of these peaks can be attributed to microwave-induced effects, as reported in Ref.~\cite{PhysRevB.110.064511}. These effects most likely originate from radiative coupling between the device and the cylindrical cavity mode, or from spurious resonances within the cryostat's low-frequency measurement lines.

\begin{figure}[H]
\centering
\includegraphics[width=0.95\textwidth]{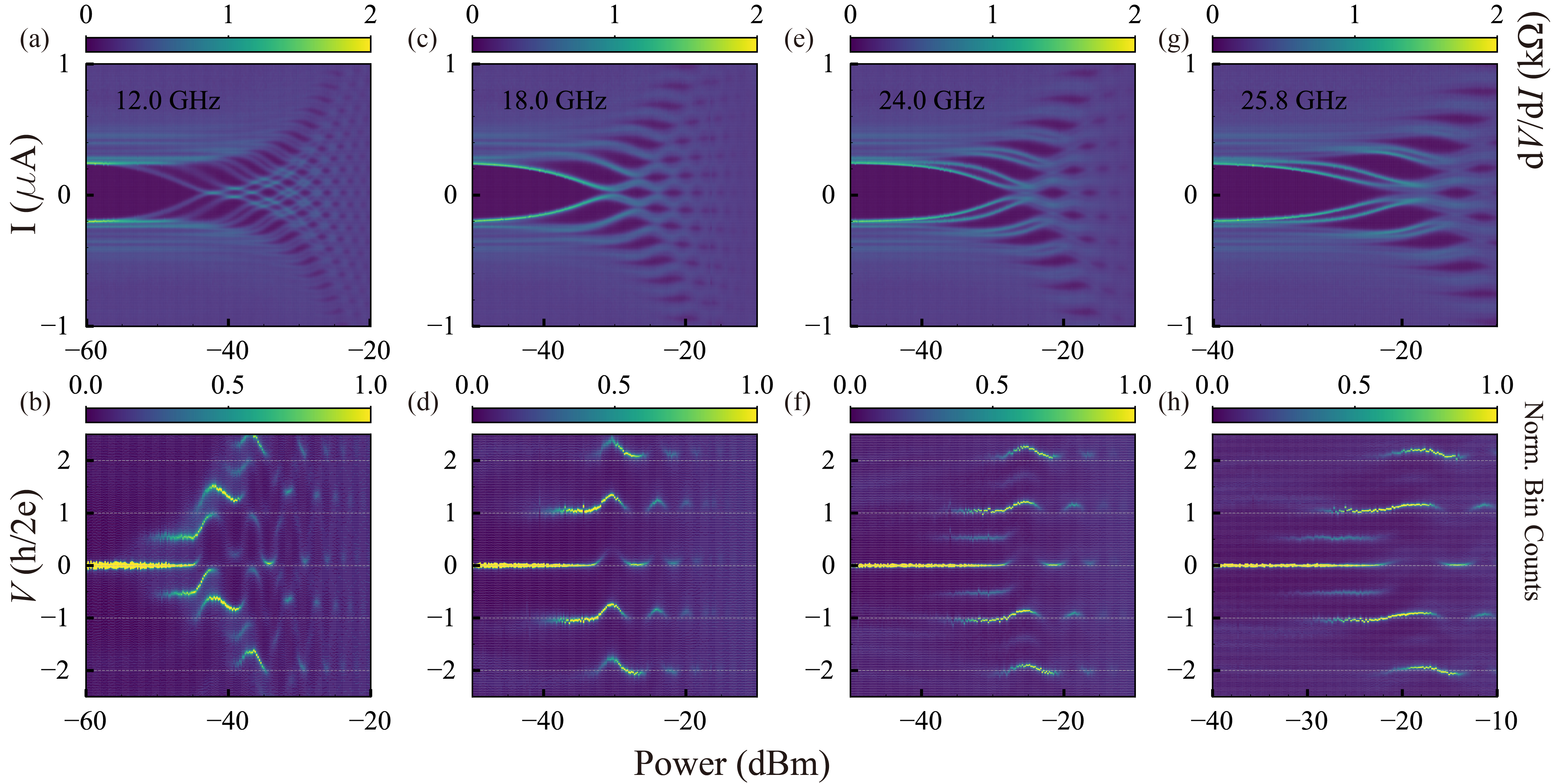}% Here is how to import EPS art
\caption{\label{fig:FigS_RF1}
Radio frequency response of JJA. Differential resistance as a function of bias current and RF excitation power at frequencies $f_{RF}$ = 12,
18, 24 and 25.8 GHz, respectively.}
\end{figure}

\subsection{Shapiro steps under radio-frequency (RF) irradiation}

% Fractional AC Josephson effect in a topological insulator proximitized by a self-formed superconductor~\cite{PhysRevB.110.064511}.

% Zero Crossing Steps and Anomalous Shapiro Maps in Graphene Josephson Junctions~\cite{Larson2020NanoLetters}.

% Josephson detection of time-reversal symmetry broken superconductivity in SnTe nanowires~\cite{Trimble2021timereversalsymmetrySnTe}.

% Anomalous phase dynamics of driven graphene Josephson junctions~\cite{PhysRevResearch.2.023093}.

% ac Josephson effect in a gate-tunable ${\mathrm{Cd}}_{3}{\mathrm{As}}_{2}$ nanowire superconducting weak link~\cite{acJosephsoneffectPhysRevB.108.094514}.

% Model for missing Shapiro steps due to bias-dependent resistance~\cite{ModelformissingShapirostepsMudi2021}.

% Asymmetric nanowire SQUID: Linear current-phase relation, stochastic switching, and symmetries~\cite{MurphyPhysRevB.96.094507}.

% Impact of Kinetic Inductance on the Critical-Current Oscillations of Nanobridge SQUIDs~\cite{PhysRevApplied.16.024013KineticInductance}.

% Limitations of the Current–Phase Relation Measurements by an Asymmetric dc - SQUID~\cite{Babich2023LimitationsoftheCurrentPhaseRelation}.

% Current–Phase Relation of a WTe${_2}$ Josephson Junction~\cite{Endres2023Current–PhaseRelation WTe}.

% 4pi-periodic Josephson supercurrent in HgTe-based topological Josephson junctions~\cite{wiedenmann20164pi}.

We experimentally observed the appearance of Shapiro steps in a Josephson junction (JJA) under radio-frequency (RF) irradiation, as shown in Fig.~\ref{fig:FigS_RF1}. RF excitation was applied via an open-ended coaxial line to the sample. The upper column of Fig.~\ref{fig:FigS_RF1} shows the evolution of differential resistance as a function of RF excitation power at different RF frequencies. The development of Shapiro steps is clearly demonstrated by the histograms in the lower column of Fig.~\ref{fig:FigS_RF1}, formed by grouping data points (equally spaced in DC current) into DC voltage bins, where Shapiro steps appear as bright streaks. 

\begin{figure}[H]
\centering
\includegraphics[width=0.7\textwidth]{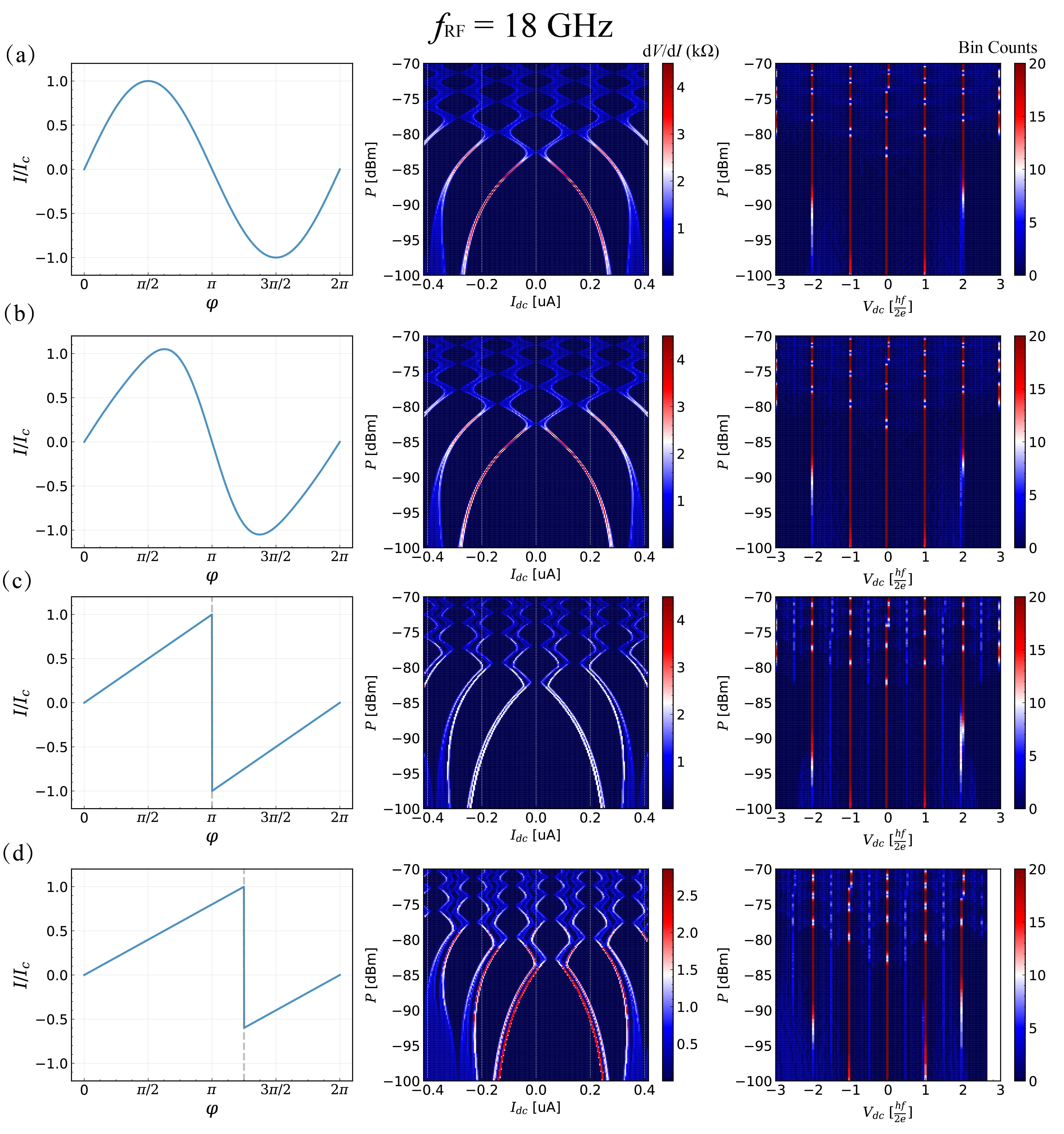}% Here is how to import EPS art
\caption{\label{fig:shapiro_sum_18GHz}
Simulated Shapiro step patterns under 18 GHz radio-frequency (RF) irradiation for different current-phase relations (CPRs). (a) Conventional sinusoidal CPR \( I(\phi) = I_c \sin(\phi) \). (b) Slightly skewed CPR with higher harmonic terms: \( I(\phi) = I_c[\sin(\phi) - 0.2\sin(2\phi) + 0.04\sin(3\phi)] \). (c) Maximally skewed sawtooth CPR. (d) Linear CPR \( I(\phi) = I_c \frac{(\phi - \phi_c \mod 2\pi) + \phi_c - 2\pi}{\phi_c} \) with \( \phi_c = 1.25\pi \). Deviations from the sinusoidal CPR enhance fractional Shapiro steps (e.g., \( n = 1/2, 3/2 \)), while the linear CPR induces asymmetric hysteresis (panel d), inconsistent with experimental observations. The slightly skewed CPR (b) best reproduces the fractional step strength observed in Fig.~\ref{fig:FigS_RF1}.  
}
\end{figure}

To simulate the dynamics of the Josephson junction, we employed the the resistively and capacitively shunted junction (RCSJ) model~\cite{PhysRevB.110.064511,Larson2020NanoLetters,Trimble2021timereversalsymmetrySnTe,PhysRevResearch.2.023093,acJosephsoneffectPhysRevB.108.094514,ModelformissingShapirostepsMudi2021,wiedenmann20164pi}, which describes the temporal evolution of the superconducting phase $\varphi (t)$ through the differential equation,
\begin{equation}
    \frac{\hbar}{2e} C \frac{d^2 \varphi}{dt^2} + \frac{\hbar}{2e} \frac{1}{R} \frac{d \varphi}{dt} + I_c f(\varphi) = I(t)
\end{equation}
where \( C \) is the junction capacitance, \( R \) is the parallel resistance, \( f(\varphi) \) describes the dimensionless current-phase relation, \( I_c \) is the critical current, and \( I(t) = I_{\text{dc}} + I_{\text{ac}} \sin(2 \pi f_RF t) \) is the bias current. By introducing normalized units \( i = I/I_c \), \( i_F = I_F/I_c \), and \( \tau = t/t_c \) with \( t_c = \left( \frac{2e}{\hbar} I_c R \right)^{-1} \), the equation transforms to  
\begin{equation}
\frac{1}{\beta} \frac{d^2 \varphi}{d\tau^2} + \frac{d \varphi}{d\tau} + f(\varphi) - i(\tau) = 0,
\end{equation}
where the Stewart-McCumber parameter \( \beta = \frac{2e}{\hbar} I_c R^2 C \) characterizes the junction dynamics.  

\begin{figure}[H]
\centering
\includegraphics[width=0.7\textwidth]{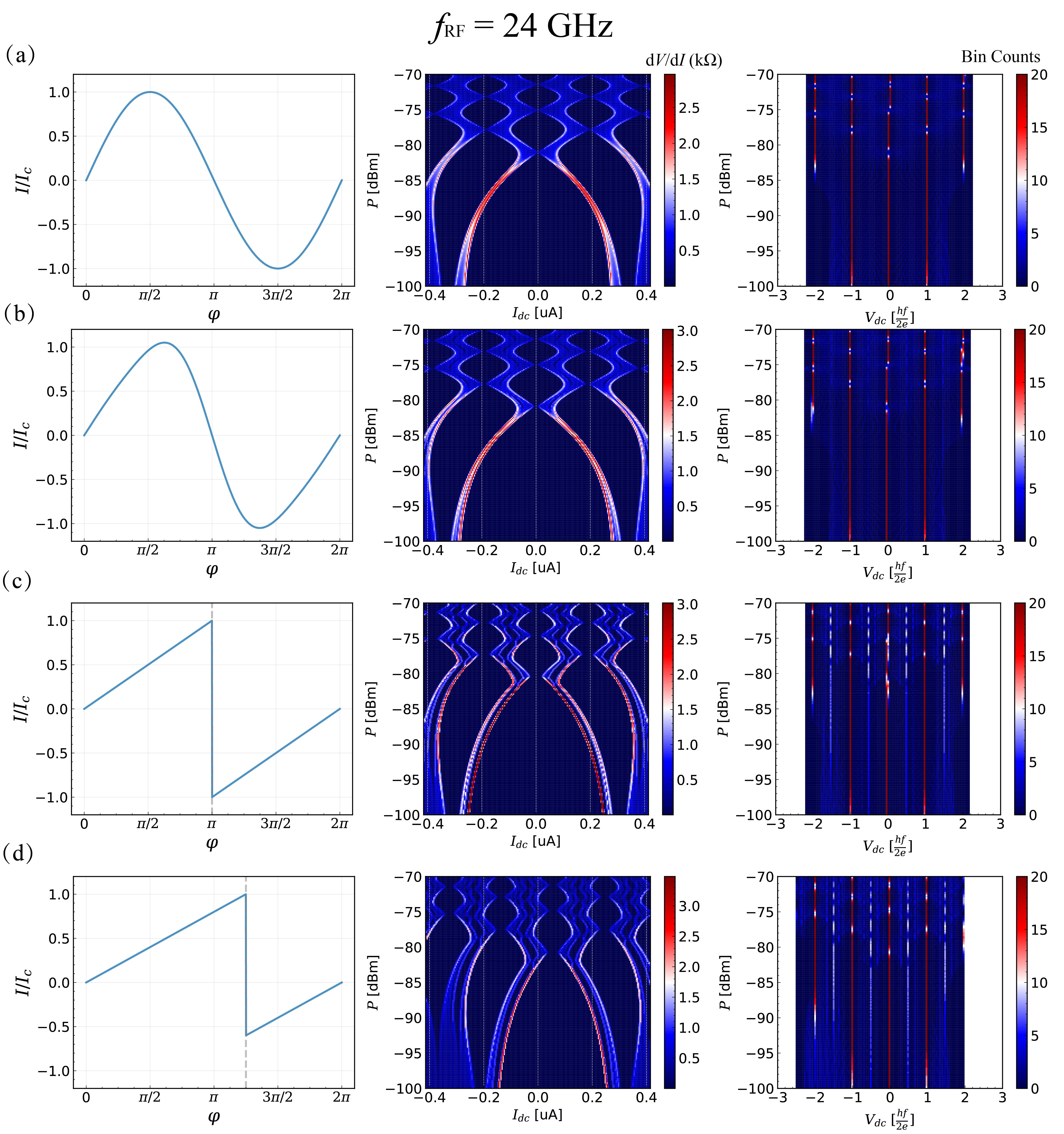}% Here is how to import EPS art
\caption{\label{fig:shapiro_sum_24GHz}
Simulated Shapiro step patterns under 24 GHz radio-frequency (RF) irradiation for different current-phase relations (CPRs).}
\end{figure}

To fit the experimental IV curve shown in Fig.~\ref{fig:FigS_IV_hysteresis}(a), we incorporated thermal noise \( I_F(t) \) through a stochastic current term \( I_F(t) \)~\cite{acJosephsoneffectPhysRevB.108.094514}. This noise term satisfies \( \langle I_F(t) \rangle = 0 \) and \( \langle I_F(t) I_F(t') \rangle = \frac{2 k_B T}{R} \delta(t - t') \). The fit result is show in Fig.~\ref{fig:FigS_IV_hysteresis}(c), achieving good agreement between numerical simulations and experimental results. The non-hysteretic current-voltage relationship in Fig.~\ref{fig:FigS_IV_hysteresis}(b) indicates that Joule overheating does not limit the retrapping current. Nevertheless, the extracted Stewart-McCumber parameter \( \beta_C \sim 1.3 \) renders the junction operates in the underdamped rather than overdamped regime.

To efficiently simulate Shapiro step patterns, we neglected capacitance (\( C = 0 \)), reducing the evolution equation to a first-order differential form. The resulting equation was solved using the SciPy odeint solver in Python, with the instantaneous voltage calculated as \( V = \hbar \dot{\phi}/2e \) at each time step. Simulations employed approximately 200,000 time steps over a range of 2000 \( t_c \), leveraging the Cython package to enhance computational speed. The DC voltage was determined by averaging the instantaneous voltage, and the DC bias applied to the device was represented in power units due to the unknown effective impedance at microwave frequencies,
\begin{equation}
I_{\text{DC}} = \left( 50 \, \Omega \times 1 \, \text{mW} \times 10^{\frac{P_{\text{RF}}}{10}} \right)^{1/2}.
\end{equation}
where \( P_{\text{RF}} \) is the power in decibels referenced to 1 mW given a \( 50 \, \Omega \) impedance.

Note that the experiment in Fig~\ref{fig:FigS_RF1} shows fractional Shapiro steps at certain RF frequencies. For comparison, our simulations considered CPRs with varying degrees of skewness, as shown in Fig.~\ref{fig:shapiro_sum_18GHz} and Fig.~\ref{fig:shapiro_sum_24GHz}. We examined four types of CPRs: a standard sinusoidal CPR (panel a), a slightly skewed CPR~\cite{Larson2020NanoLetters}   \( I(\phi) = I_c[\sin(\phi) - 0.2\sin(2\phi) + 0.04\sin(3\phi)] \) (panel b), a maximally skewed sawtooth CPR (panel c), and a linear CPR~\cite{Babich2023LimitationsoftheCurrentPhaseRelation} \( I_c \frac{(\varphi - \varphi_c \mod 2\pi) + \varphi_c - 2\pi}{\varphi_c} \) with an arbitrarily chosen \( \varphi_c = 1.25\pi \) (panel d). The simulation results for RF frequencies of 18 GHz and 24 GHz are presented in Fig.~\ref{fig:shapiro_sum_18GHz} and Fig.~\ref{fig:shapiro_sum_24GHz}, respectively.

Our results indicate that deviations from the standard sinusoidal CPR can lead to the appearance of fractional Shapiro steps. Notably, the maximally skewed sawtooth CPR produces more pronounced fractional steps, while the linear CPR with an arbitrary \( \phi_c \) introduces asymmetry in the Shapiro steps, resulting in hysteresis behavior as seen in Fig.~\ref{fig:shapiro_sum_18GHz}(d) and Fig.~\ref{fig:shapiro_sum_24GHz}(d). Critically, such hysteresis and asymmetry are absent in the experimental data.  Additionally, in the context of the Little-Parks (LP) diamond mechanism~\cite{MurphyPhysRevB.96.094507,PhysRevApplied.16.024013KineticInductance}, overlap occurs only if \( \phi_c > \pi \). Given that our experimental shaprio results do not exhibit distinct asymmetry, we can ruled out the possibility of a linear CPR.
Comparing our simulations with the experimental data, we find that the slightly skewed CPR most accurately reproduces the fractional Shapiro steps observed in the experiment. This suggests that the experimental results can be attributed to a slight deviation from the standard sinusoidal CPR.

% This conversion framework, central to high-frequency engineering, reconciles logarithmic power metrics with linear circuit parameters: starting from \( P_{\text{dBm}} = 10 \log_{10}(P / 1 \, \text{mW}) \), power in watts is \( P = 10^{(P_{\text{dBm}} - 30)/10} \), from which RMS voltage \( V_{\text{rms}} = \sqrt{PZ_0} \) and peak voltage \( V_{\text{peak}} = V_{\text{rms}}\sqrt{2} \) (for sinusoidal signals) are derived, leading to peak current \( I_{\text{peak}} = \sqrt{2P/Z_0} \). This ensures rigorous signal integrity in RF systems, critical for microwave design and quantum device control such as Josephson junctions, by systematically converting decibel-referenced power to amplitude parameters while avoiding unphysical artifacts through computational safeguards.

% 1. **dBm to Power (W):**  
%   \( P = 10^{(P_{\text{dBm}} - 30)/10} \), where \( P_{\text{dBm}} \) is the power in dBm (relative to 1 mW).

% 2. **Power to RMS Voltage:**  
%   \( V_{\text{rms}} = \sqrt{P \cdot Z_0} \), with \( Z_0 \) as the impedance (e.g., 50 Ω).

% 3. **RMS to Peak Voltage:**  
%   \( V_{\text{peak}} = V_{\text{rms}} \cdot \sqrt{2} \), assuming a sinusoidal waveform.

% 4. **Peak Voltage to Peak Current:**  
%   \( I_{\text{peak}} = V_{\text{peak}} / Z_0 \), derived from Ohm’s Law.

\subsection{Similar interference pattern  observed in JJB}

\begin{figure}[H]
\centering
\includegraphics[width=0.75\textwidth]{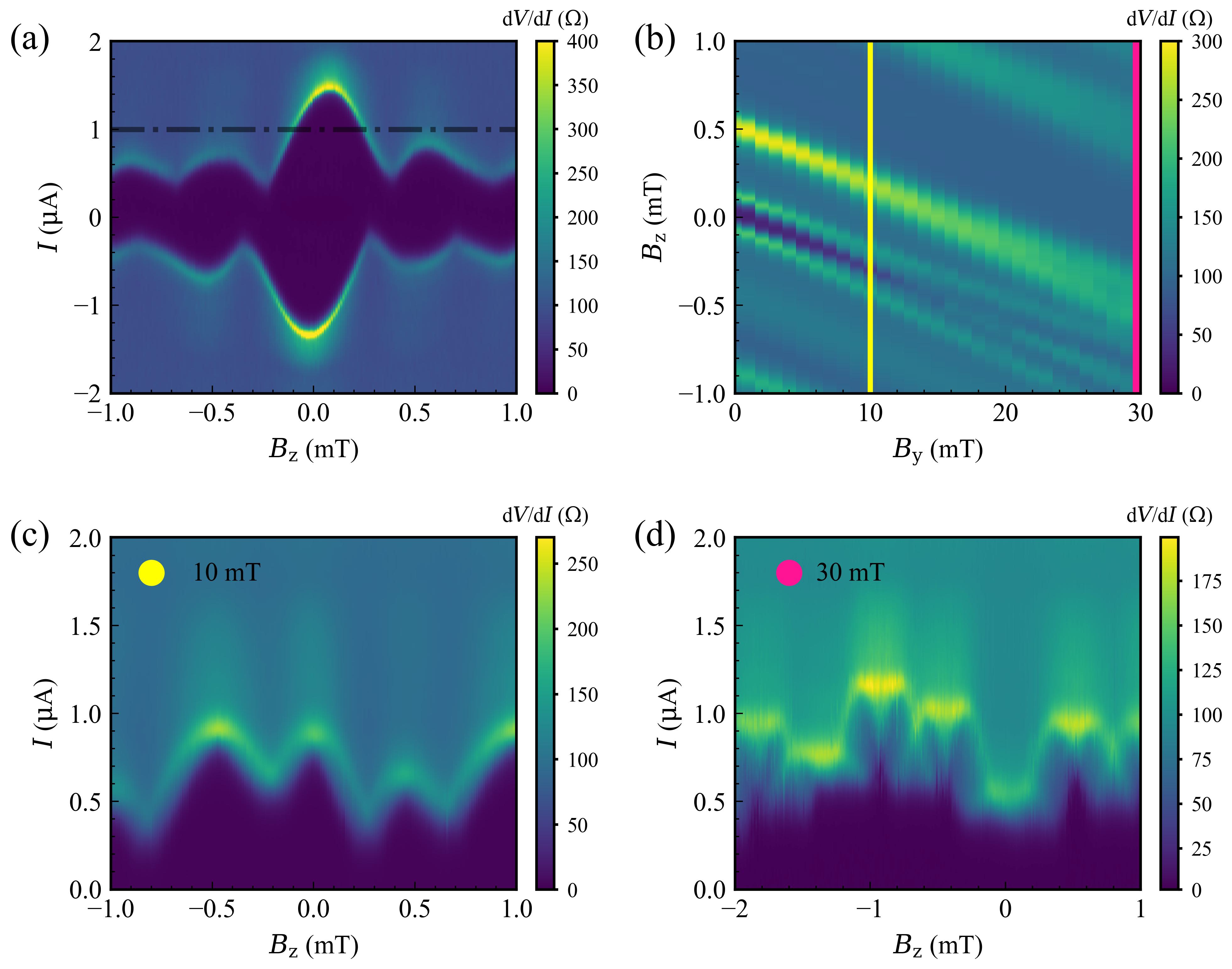}% Here is how to import EPS art
\caption{\label{fig:FigS_J1}
Evolution of the interference pattern of JJB. (a) Differential resistance map versus $I$ and $B_z$, exhibits a Fraunhofer-like  pattern, revealing a relatively uniform current distribution in JJB. (b) Evolution of the interference pattern  versus in-plane magnetic field $B_y$. Note that the map is obtained  with  a finite  bias current ($I= 1 \; \mathrm{\mu A}$) as indicated by the black  dot-dashed line in (a), differing from  $I = 0$ in the case of JJA. (c) and (d) Interference spectra under $B_y = 10, 30 \; \mathrm{mT}$,  respectively, corresponding to the solid lines indicated in (b).}
\end{figure}

An analogous evolution of the interference pattern is observed in the Josephson junction-based (JJB) device (Fig.~\ref{fig:FigS_J1}). Under in-plane magnetic fields, the critical supercurrent is suppressed, accompanied by a transition from a Fraunhofer-like diffraction pattern to a superconducting quantum interference device (SQUID)-like modulation, as shown in Fig.~\ref{fig:FigS_J1}(c).

A similar phenomenon has been demonstrated in planar graphene Josephson junctions~\cite{PlanargrapheneNbSeJosephsonjunctionsPhysRevB.103.115401}, where the orbital contribution from interfacial ripple structures plays a dominant role. In our system, the potential Zeeman splitting effect has been experimentally ruled out through systematic analysis of the weak in-plane magnetic field magnitude. We therefore propose that our experimental findings likely attributed to the orbital effect arising from intrinsic ripple structures - an inherent characteristic unavoidably present in weakly-coupled Josephson junctions.

%\appendix
%\input{appendixes}

% \bibliography{ref_SM}% Produces the bibliography via BibTeX.
